\documentclass[pra,print,showpacs,superscriptaddress,twocolumn]{revtex4}
\usepackage{amssymb}
\usepackage{amsmath}
\usepackage{epsfig}
\usepackage{color}
\usepackage{graphics, graphicx}
\usepackage{bbold}
\usepackage{psfrag}
\usepackage{mathcomp}
\usepackage{subfigure}
\usepackage{verbatim}
\usepackage{color}
\usepackage[colorlinks,citecolor=blue]{hyperref}
\usepackage{dcolumn}
\usepackage{bm}
\usepackage{bbm}
\usepackage{mathrsfs}
\usepackage{amsfonts}
\usepackage{subfigure}

\begin{document}

\title{Quantum phases of a two-dimensional polarized degenerate Fermi gas in
an optical cavity}
\author{Yanlin Feng}
\affiliation{State Key Laboratory of Quantum Optics and Quantum Optics Devices, Institute
of Laser Spectroscopy, Shanxi University, Taiyuan, Shanxi 030006, China}
\affiliation{Collaborative Innovation Center of Extreme Optics, Shanxi University,
Taiyuan, Shanxi 030006, China}
\author{Kuang Zhang}
\affiliation{State Key Laboratory of Quantum Optics and Quantum Optics Devices, Institute
of Laser Spectroscopy, Shanxi University, Taiyuan, Shanxi 030006, China}
\affiliation{Collaborative Innovation Center of Extreme Optics, Shanxi University,
Taiyuan, Shanxi 030006, China}
\author{Jingtao Fan}
\affiliation{State Key Laboratory of Quantum Optics and Quantum Optics Devices, Institute
of Laser Spectroscopy, Shanxi University, Taiyuan, Shanxi 030006, China}
\affiliation{Collaborative Innovation Center of Extreme Optics, Shanxi University,
Taiyuan, Shanxi 030006, China}
\author{Feng Mei}
\thanks{raulmei@163.com}
\affiliation{State Key Laboratory of Quantum Optics and Quantum Optics Devices, Institute
of Laser Spectroscopy, Shanxi University, Taiyuan, Shanxi 030006, China}
\affiliation{Collaborative Innovation Center of Extreme Optics, Shanxi University,
Taiyuan, Shanxi 030006, China}
\author{Gang Chen}
\thanks{chengang971@163.com}
\affiliation{State Key Laboratory of Quantum Optics and Quantum Optics Devices, Institute
of Laser Spectroscopy, Shanxi University, Taiyuan, Shanxi 030006, China}
\affiliation{Collaborative Innovation Center of Extreme Optics, Shanxi University,
Taiyuan, Shanxi 030006, China}
\author{Suotang Jia}
\affiliation{State Key Laboratory of Quantum Optics and Quantum Optics Devices, Institute
of Laser Spectroscopy, Shanxi University, Taiyuan, Shanxi 030006, China}
\affiliation{Collaborative Innovation Center of Extreme Optics, Shanxi University,
Taiyuan, Shanxi 030006, China}

\begin{abstract}
In this paper we analytically investigate the ground-state properties of a
two-dimensional polarized degenerate Fermi gas in a high-finesse optical
cavity, which is governed by a generalized Fermi-Dicke model with tunable
parameters. By solving the photon-number dependent Bogoliubov--de--Gennes
equation, we find rich quantum phases and phase diagrams, which depend
crucially on the fermion-photon coupling strength, the fermion-fermion
interaction strength, and the atomic resonant frequency (effective Zeeman
field). In particular, without the fermion-fermion interaction and with a
weak atomic resonant frequency, we find a mixed phase that the normal phase
with two Fermi surfaces and the superradiant phase coexist, and reveal a
first-order phase transition from this normal phase to the superradiant
phase. With the intermediate fermion-fermion interaction and fermion-photon
coupling strengths, we predict another mixed phase that the superfluid and
superradiant phases coexist. Finally, we address briefly how to detect these
predicted quantum phases and phase diagrams in experiments.
\end{abstract}

\maketitle

\section{Introduction}

The experimental combination of a Bose-Einstein condensate with a
high-finesse optical cavity \cite{FB07,YC07} opens a conceptually new regime
of both cavity quantum electrodynamics and ultracold atoms. In this
combination, all ultracold bosons, occupying the same quantum state,
interact identically with a single-mode quantized field, and thus, a strong
collective matter-field interaction can be achieved. Moreover, cavities can
generate unconventional dynamical optical potentials, which induce rich
nonequilibrium and strongly-corrected many-body phenomena \cite{HR13}. For
example, when pumped transversely, the spinless ultracold bosons in the
cavity-induced dynamical optical potentials undergo self-organization \cite%
{PD02,DN06,DN08}, which has been observed experimentally \cite{KB10,KB11}
and has been regarded as an equivalence to the well-known superradiant (SR)
phase transition in an effective Dicke model \cite{DH54}.

Motivated by near-term experimental prospects, another fundamental
interaction between ultracold fermions and a high-finesse optical cavity has
been investigated theoretically. Since at lower temperature fermions exhibit
quite different behavior than bosons, exotic physics is expected to arise in
this new platform \cite%
{RK10,QS11,MM12,XG12,BP13,JK14,FP14,YC14,JP14,YC15,CK16,AS16,ASF16,WZ16}. In
particular, followed by the experimental scheme in Ref.~\cite{KB10,KB11},
three groups have considered simultaneously spinless fermions in the
cavity-induced dynamical optical potential \cite{JK14,FP14,YC14}. They have
found that the Fermi statistics plays a dominate role in the SR phase
transition at moderate and high densities. At the moderate density, the
Fermi surface displays a nesting structure and strongly enhances
superradiance, which is, however, suppressed largely at high density, due to
the Pauli blocking effect. In addition, by introducing a cavity-assisted
spin-orbit coupling \cite{YD14,LD14}, a topological SR phase has been
predicted \cite{JP14}. Recently, the cavity-induced artificial magnetic
field \cite{CK16}, chiral phases \cite{AS16}, and non-trivial topological
states \cite{ASF16} have been created. Moreover, when fermions are gauge
coupled to a cavity mode, a SR phase with an infinitesimal pumping
threshold, which induces a directed particle flow, has been found for an
infinite lattice \cite{WZ16}.

In this paper, followed by the experimental scheme in Ref.~\cite{KJA12,MPB14}%
, we consider a two-dimensional (2D) polarized degenerate Fermi gas in a
high-finesse optical cavity. When introducing two Raman transitions induced
by the quantized cavity field and two transverse pumping lasers, we first
realize a generalized Fermi-Dicke model, in which all parameters, including
the fermion-photon coupling strength, the fermion-fermion interaction
strength, and the atomic resonant frequency (effective Zeeman field), can be
controlled independently. Then, based on a photon-number dependent
Bogoliubov--de--Gennes (BdG) equation, we reveal rich quantum phases and
phase diagrams, which depend crucially on these tunable parameters. In
particular, without the fermion-fermion interaction and with a weak atomic
resonant frequency, we find a mixed phase that the normal phase with two
Fermi surfaces and the SR phase coexist, and reveal a first-order phase
transition from this normal phase to the SR phase. With the intermediate
fermion-fermion interaction and fermion-photon coupling strengths, we
predict another mixed phase that the superfluid (SF) and SR phases coexist.
Finally, we address briefly how to detect the predicted quantum phases and
phase diagrams in experiments.

This paper is organized as follows. In Sec.~\ref{Mode and Hamiltonian}, we
present an experimentally-feasible scheme to realize a generalized
Fermi-Dicke model with tunable parameters. In Sec.~\ref{Ground-state}, we
derive a photon-number dependent BdG equation, and then obtain the
ground-state energy and the mean-field gap, particle number, and SR
equations. In Secs.~\ref{Phase I} and \ref{Phase II}, we reveal rich quantum
phases and phase diagrams without or with the fermion-fermion two-body
interaction, respectively. The parameter estimation and possible
experimental observation are addressed in Sec.~\ref{Parameter estimation},
and the brief discussion and conclusion are given in Sec.~\ref{Discussion}.%
\newline

\section{Model and Hamiltonian}

\label{Mode and Hamiltonian}

\begin{figure}[t]
\centering
\includegraphics[width=5.5cm]{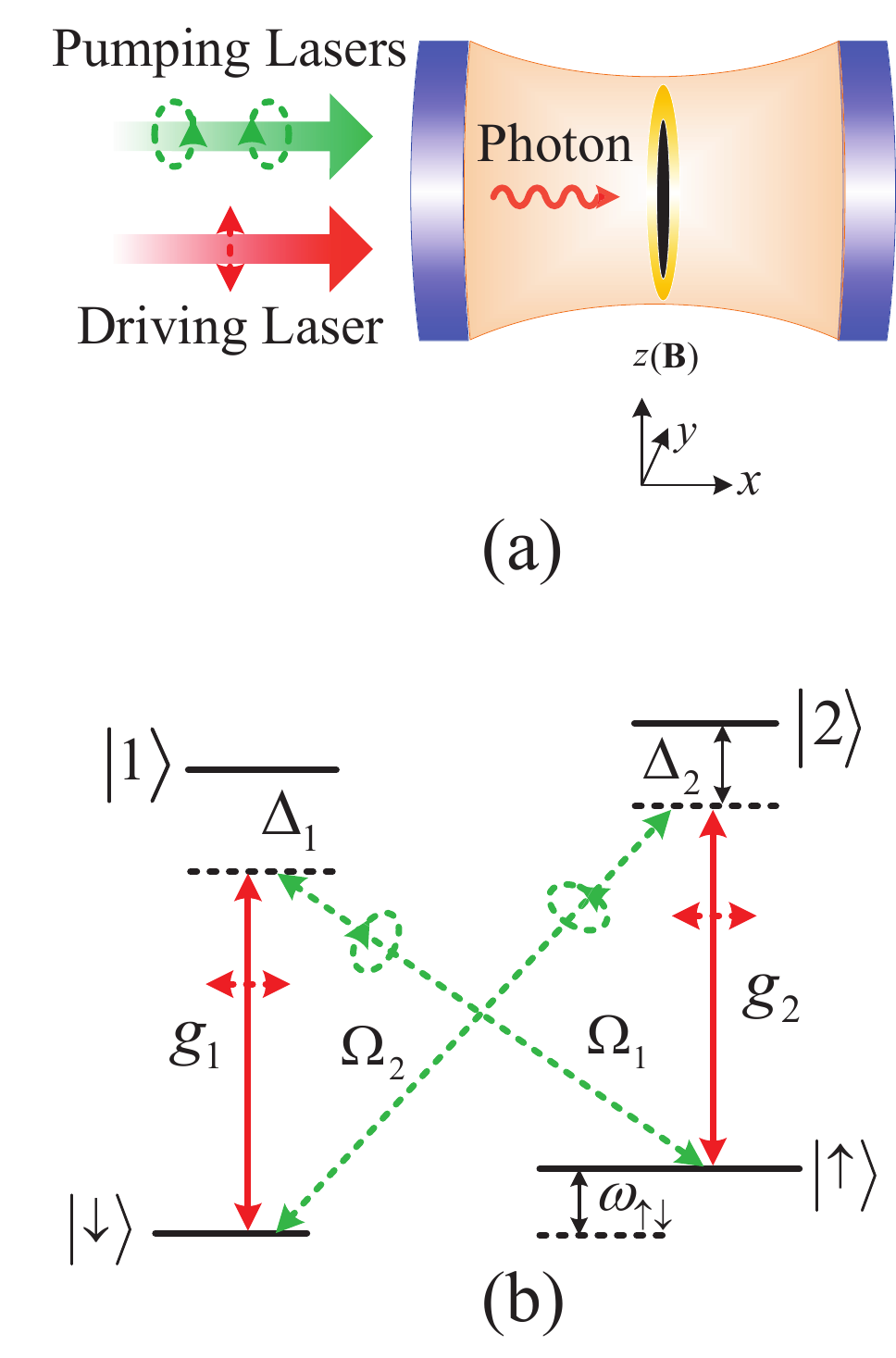}\newline
\caption{(a) Our proposed schematic setup that all ultracold fermions (black
online), which are confined in a far-of-resonance optical trap (yellow
online) of the $yz$ plane by a tightly-radial confinement along the $x$
direction, are coupled with a high-finesse optical cavity. The cavity mode
is driven by a linearly-polarized driving laser (with frequency $\protect%
\omega _{l}$), which propagates along the $x$ direction. Two transverse
pumping lasers (with frequencies $\protect\omega _{A}$ and $\protect\omega %
_{B}$), which are left- and right-handed circular polarized in the $yz$
plane, propagate along the $x$ direction and induce two Raman processes by
combining the quantized cavity field. In order to obtain a time-independent
Hamiltonian, these frequencies should satisfy the following condition: $%
\protect\omega _{l}=\left( \protect\omega _{A}+\protect\omega _{B}\right) /2$%
; see the detailed derivation in the main text. The magnetic field $\mathbf{B%
}$ is applied along the positive $z$ direction and produces a Zeeman shift
between two hyperfine ground states. (b) The atomic energy levels and their
transitions. Each fermion has two ground states ($\left\vert \uparrow
\right\rangle $ and $\left\vert \downarrow \right\rangle $) and two excited
states ($\left\vert 1\right\rangle $ and $\left\vert 2\right\rangle $). The $%
\left\vert \downarrow \right\rangle $ $\longleftrightarrow \left\vert
1\right\rangle $ and $\left\vert \uparrow \right\rangle $ $%
\longleftrightarrow \left\vert 2\right\rangle $ transitions (red solid
lines) are caused by the quantized cavity field with fermion-photon coupling
strengths $g_{1}$ and $g_{2}$. The $\left\vert \uparrow \right\rangle $ $%
\longleftrightarrow \left\vert 1\right\rangle $ and $\left\vert \downarrow
\right\rangle $ $\longleftrightarrow \left\vert 2\right\rangle $ transitions
(green dashed lines) are governed by the transverse pumping lasers with Rabi
frequencies $\Omega _{1}$ and $\Omega _{2}$. $\protect\omega _{\uparrow
\downarrow }=\protect\omega _{\uparrow }-\protect\omega _{\downarrow }$ is
the resonant frequency between the ground states $\left\vert \uparrow
\right\rangle $ and $\left\vert \downarrow \right\rangle $ with
eigenfrequencies $\protect\omega _{\uparrow }$ and $\protect\omega %
_{\downarrow }$. $\Delta _{1}$ and $\Delta _{2}$ are the detunings from the
excited states $\left\vert 1\right\rangle $ and $\left\vert 2\right\rangle $%
. }
\label{fig1}
\end{figure}

Figure \ref{fig1} shows our proposed scheme that all ultracold fermions are
coupled with a high-finesse optical cavity supporting a single-mode photon.
As illustrated in Fig.~\ref{fig1}(a), the fermions in the optical cavity are
confined in a far-of-resonance optical trap ($yz$\ plane) by a
tightly-radial confinement along the $x$\ direction. The cavity mode is
driven by a linearly-polarized laser and the fermions are pumped by two
transverse lasers, which are left- and right-handed circular polarized in
the $yz$\ plane. In addition, each fermion has four levels, including two
ground states ($\left\vert \uparrow \right\rangle $ and $\left\vert
\downarrow \right\rangle $) and two excited states ($\left\vert
1\right\rangle $ and $\left\vert 2\right\rangle $), as shown in Fig.~\ref%
{fig1}(b). The quantized cavity field and the two transverse pumping lasers
induce two Raman processes; see more details in the caption.

Formally, the total time-dependent 2D Hamiltonian is written as%
\begin{equation}
\hat{H}_{\text{T}}(t)=\hat{H}_{\text{F}}+\hat{H}_{\text{P}}+\hat{H}_{\text{D}%
}(t)+\hat{H}_{\text{AR}}(t)+\hat{H}_{\text{AP}}+\hat{H}_{\text{INT}}.
\label{S1}
\end{equation}%
Here, the Hamiltonian of the free four-level fermions is given by%
\begin{equation}
\hat{H}_{\text{F}}=\sum_{i=1,2,\uparrow ,\downarrow }\int d^{2}\mathbf{r}%
\hat{\psi}_{i}^{\dagger }\left( \mathbf{r}\right) \left( \frac{\mathbf{\hat{p%
}}^{2}}{2M}-\mu +\omega _{i}\right) \hat{\psi}_{i}\left( \mathbf{r}\right) ,
\label{FH}
\end{equation}%
where $\hat{\psi}_{i}^{\dagger }\left( \mathbf{r}\right) $ and $\hat{\psi}%
_{i}\left( \mathbf{r}\right) $ ($i=1,2,\downarrow ,\uparrow $) are the
creation and annihilation operators of the fermionic fields, $M$ is the atom
mass, $\mu $ is the chemical potential, and $\omega _{i}$ are the
eigenfrequencies of all quantum states. The Hamiltonian of the quantized
cavity field, together with the driving laser, is written as
\begin{equation}
\hat{H}_{\text{P}}+\hat{H}_{\text{D}}(t)=\omega _{c}\hat{a}^{\dag }\hat{a}%
+\varepsilon \left( \hat{a}e^{i\omega _{l}t}+\hat{a}^{\dagger }e^{-i\omega
_{l}t}\right) ,  \label{PE}
\end{equation}%
where $\hat{a}^{\dag }$ and $\hat{a}$ are the creation and annihilation
operators of the quantized cavity field with frequency $\omega _{c}$, and $%
\varepsilon $ ($\omega _{l}$) is the magnitude (frequency) of the driving
laser. Under the rotating-wave approximation, the Hamiltonian, which
describes the interaction between the fermionic fields and the two
transverse pumping lasers, reads
\begin{eqnarray}
\hat{H}_{\text{AR}}(t) &=&\frac{1}{2}\int d^{2}\mathbf{r}\left[ \Omega _{1}%
\hat{\psi}_{1}^{\dagger }\left( \mathbf{r}\right) \hat{\psi}_{\uparrow
}\left( \mathbf{r}\right) e^{-i\omega _{A}t}\right.  \notag \\
&&+\left. \Omega _{2}\hat{\psi}_{2}^{\dagger }\left( \mathbf{r}\right) \hat{%
\psi}_{\downarrow }\left( \mathbf{r}\right) e^{-i\omega _{B}t}+\text{H.c.}%
\right] ,  \label{S3}
\end{eqnarray}%
where $\Omega _{1}$ and $\Omega _{2}$ ($\omega _{A}$ and $\omega _{B}$) are
the Rabi frequencies (frequencies) of the transverse pumping lasers and H.c.
denotes the Hermitian conjugate, whereas the Hamiltonian for governing the
interaction between the fermionic and quantized cavity fields is given by%
\begin{equation}
\hat{H}_{\text{AP}}\!=\!\!\int \!d^{2}\mathbf{r}\left\{ \left[ g_{1}\hat{\psi%
}_{1}^{\dagger }\left( \mathbf{r}\right) \hat{\psi}_{\downarrow }\left(
\mathbf{r}\right) \!+\!g_{2}\hat{\psi}_{2}^{\dagger }\left( \mathbf{r}%
\right) \hat{\psi}_{\uparrow }\left( \mathbf{r}\right) \right] \hat{a}\!+\!%
\text{H.c.}\!\right\} ,  \label{S4}
\end{equation}%
where $g_{1}$ and $g_{2}$ are both the fermion-photon coupling strengths. In
addition, here we only consider the attractive contact interaction between
the ground states $\left\vert \uparrow \right\rangle $ and $\left\vert
\downarrow \right\rangle $ since the excited states are eliminated
adiabatically, as will be shown below. Therefore, the two-body interacting
Hamiltonian is given by%
\begin{equation}
\hat{H}_{\text{INT}}=\lambda \int d^{2}\mathbf{r}\hat{\psi}_{\uparrow
}^{\dagger }\left( \mathbf{r}\right) \hat{\psi}_{\downarrow }^{\dagger
}\left( \mathbf{r}\right) \hat{\psi}_{\downarrow }\left( \mathbf{r}\right)
\hat{\psi}_{\uparrow }\left( \mathbf{r}\right) ,  \label{S5}
\end{equation}%
where $\lambda $ is the negative interaction strength, i.e., $\lambda \;<\;0$%
.

For the time-dependent Hamiltonian (\ref{S1}), we first perform a unitary
transformation $\hat{U}(t)=\exp \left( i\hat{H}^{\prime }t\right) $, where
\begin{eqnarray}
\hat{H}^{\prime } &=&\omega _{l}\hat{a}^{\dag }\hat{a}+\frac{\omega _{B}}{2}%
\left[ \hat{\psi}_{2}^{\dagger }\left( \mathbf{r}\right) \hat{\psi}%
_{2}\left( \mathbf{r}\right) -\hat{\psi}_{\downarrow }^{\dagger }\left(
\mathbf{r}\right) \hat{\psi}_{\downarrow }\left( \mathbf{r}\right) \right]
\notag \\
&&+\frac{\omega _{A}}{2}\left[ \hat{\psi}_{1}^{\dagger }\left( \mathbf{r}%
\right) \hat{\psi}_{1}\left( \mathbf{r}\right) -\hat{\psi}_{\uparrow
}^{\dagger }\left( \mathbf{r}\right) \hat{\psi}_{\uparrow }\left( \mathbf{r}%
\right) \right]  \label{UNT}
\end{eqnarray}%
with $\omega _{l}=\left( \omega _{B}+\omega _{A}\right) /2$, to obtain a
time-independent Hamiltonian $\hat{H}_{1}=\hat{U}(t)\hat{H}_{\text{T}}(t)%
\hat{U}^{\dagger }(t)+i\left[ \partial \hat{U}(t)/\partial t\right] \hat{U}%
^{\dagger }(t)$, i.e.,%
\begin{widetext}
\begin{eqnarray}
\hat{H}_{1} &=&\tilde{\omega}\hat{a}^{\dag }\hat{a}+\varepsilon \left( \hat{a%
}+\hat{a}^{\dagger }\right) +\sum_{i=1,2,\uparrow ,\downarrow }\int d^{2}%
\mathbf{r}\hat{\psi}_{i}^{\dagger }\left( \mathbf{r}\right) \left( \frac{%
\mathbf{\hat{p}}^{2}}{2M}-\mu \right) \hat{\psi}_{i}\left( \mathbf{r}\right) +\int
d^{2}\mathbf{r}\left[ \Delta _{1}\hat{\psi}_{1}^{\dagger }\left( \mathbf{r}%
\right) \hat{\psi}_{1}\left( \mathbf{r}\right) +\Delta _{2}\hat{\psi}%
_{2}^{\dagger }\left( \mathbf{r}\right) \hat{\psi}_{2}\left( \mathbf{r}%
\right) \right]   \notag \\
&&+\int d^{2}\mathbf{r}\left[ \tilde{\omega}_{\uparrow }\hat{\psi}_{\uparrow
}^{\dagger }\left( \mathbf{r}\right) \hat{\psi}_{\uparrow }\left( \mathbf{r}%
\right) +\tilde{\omega}_{\downarrow }\hat{\psi}_{\downarrow }^{\dagger
}\left( \mathbf{r}\right) \hat{\psi}_{\downarrow }\left( \mathbf{r}\right) %
\right] +\frac{1}{2}\int d^{2}\mathbf{r}\left[ \Omega _{1}\hat{\psi}%
_{1}^{\dagger }\left( \mathbf{r}\right) \hat{\psi}_{\uparrow }\left( \mathbf{%
r}\right) +\Omega _{2}\hat{\psi}_{2}^{\dagger }\left( \mathbf{r}\right) \hat{%
\psi}_{\downarrow }\left( \mathbf{r}\right) +\text{H.c.}\right]   \notag \\
&&+\lambda \int d^{2}\mathbf{r}\hat{\psi}_{\uparrow }^{\dagger }\left(
\mathbf{r}\right) \hat{\psi}_{\downarrow }^{\dagger }\left( \mathbf{r}%
\right) \hat{\psi}_{\downarrow }\left( \mathbf{r}\right) \hat{\psi}%
_{\uparrow }\left( \mathbf{r}\right) +\int d^{2}\mathbf{r}\left\{ \left[
g_{1}\hat{\psi}_{1}^{\dagger }\left( \mathbf{r}\right) \hat{\psi}%
_{\downarrow }\left( \mathbf{r}\right) +g_{2}\hat{\psi}_{2}^{\dagger }\left(
\mathbf{r}\right) \hat{\psi}_{\uparrow }\left( \mathbf{r}\right) \right]
\hat{a}+\text{H.c.}\right\}.  \label{S8}
\end{eqnarray}%
\end{widetext}where $\tilde{\omega}=\omega _{c}-\omega _{l}$ is the
effective cavity frequency, $\Delta _{1}=\omega _{1}-\omega _{A}/2$ ($\Delta
_{2}=\omega _{2}-\omega _{B}/2$) is the detuning from the excited state $%
\left\vert 1\right\rangle $ ($\left\vert 2\right\rangle $), and $\tilde{%
\omega}_{\uparrow }=$ $\omega _{\uparrow }+\omega _{A}/2$ ($\tilde{\omega}%
_{\downarrow }=\omega _{\downarrow }+\omega _{B}/2$) is the effective
eigenfrequency of the ground state $\left\vert \uparrow \right\rangle $ ($%
\left\vert \downarrow \right\rangle $).

In experiments \cite{KB10,KB11,MPB14}, a weak driving ($\varepsilon
\rightarrow 0$) and large detunings ($\left\vert \Delta _{1,2}\right\vert
\gg \left\{ \Omega _{1,2},g_{1,2},\tilde{\omega},\omega _{0}\right\} $) are
usually taken into account. In such case, the term $\varepsilon \left(
a+a^{\dagger }\right) $ in the Hamiltonian (\ref{S8}) can be neglected and
both the excited states $\left\vert 1\right\rangle $ and $\left\vert
2\right\rangle $ can be eliminated adiabatically \cite{FD07,JF14}.
Therefore, we obtain
\begin{widetext}
\begin{eqnarray}
\hat{H} &=&\tilde{\omega}\hat{a}^{\dag}\hat{a}+\sum_{\sigma =\uparrow
,\downarrow}\int d^{2}\mathbf{r}\hat{\psi}_{\sigma}^{\dagger}\left(
\mathbf{r}\right) \left(\frac{\mathbf{p}^{2}}{2M}-\mu \right) \hat{\psi}%
_{\sigma}\left( \mathbf{r}\right) +\int d^{2}\mathbf{r}\left[ \tilde{\omega}%
_{\uparrow}\hat{\psi}_{\uparrow }^{\dagger }\left( \mathbf{r}\right) \hat{%
\psi}_{\uparrow}\left(\mathbf{r}\right) +\tilde{\omega}_{\downarrow }\hat{%
\psi}_{\downarrow }^{\dagger }\left( \mathbf{r}\right) \hat{\psi}%
_{\downarrow}\left( \mathbf{r}\right) \right]   \notag \\
&&+\lambda \int d^{2}\mathbf{r}\hat{\psi}_{\uparrow }^{\dagger }\left(
\mathbf{r}\right) \hat{\psi}_{\downarrow }^{\dagger }\left(\mathbf{r}%
\right) \hat{\psi}_{\downarrow }\left( \mathbf{r}\right) \hat{\psi}%
_{\uparrow}\left(\mathbf{r}\right) +\int d^{2}\mathbf{r}\left[\frac{%
|g_{2}|^{2}}{\Delta _{2}}\hat{\psi}_{\uparrow }^{\dagger}\left(\mathbf{r}%
\right) \hat{\psi}_{\uparrow }\left( \mathbf{r}\right) +\frac{|g_{1}|^{2}}{%
\Delta _{1}}\hat{\psi}_{\downarrow }^{\dagger }\left(\mathbf{r}\right) \hat{%
\psi}_{\downarrow}\left(\mathbf{r}\right) \right] \hat{a}^{\dagger }\hat{a}
\notag \\
&&+\frac{1}{2}\int d^{2}\mathbf{r}\left\{ \left[ \frac{g_{2}\Omega
_{2}^{\ast }}{\Delta _{2}}\hat{\psi}_{\downarrow }^{\dagger }\left( \mathbf{r%
}\right) \hat{\psi}_{\uparrow }\left( \mathbf{r}\right) +\frac{g_{1}\Omega
_{1}^{\ast }}{\Delta _{1}}\hat{\psi}_{\uparrow }^{\dagger }\left( \mathbf{r}%
\right) \hat{\psi}_{\downarrow }\left( \mathbf{r}\right) \right] \hat{a}+%
\text{H.c.}\right\} .  \label{S9}
\end{eqnarray}%
\end{widetext}

When the parameters are chosen as
\begin{equation}
\frac{|g_{1}|^{2}}{\Delta _{1}}=\frac{|g_{2}|^{2}}{\Delta _{2}}\text{, \ \ \
\ }\frac{g_{1}\Omega _{1}^{\ast }}{\Delta _{1}}=\frac{g_{2}\Omega _{2}^{\ast
}}{\Delta _{2}},  \label{S10}
\end{equation}%
the Hamiltonian (\ref{S9}) becomes%
\begin{eqnarray}
\hat{H} &=&\omega \hat{a}^{\dag }\hat{a}+\sum_{\sigma =\uparrow ,\downarrow
}\int d^{2}\mathbf{r}\hat{\psi}_{\sigma }^{\dagger }\left( \mathbf{r}\right)
\left( \frac{\mathbf{\hat{p}}^{2}}{2M}-\mu \right) \hat{\psi}_{\sigma
}\left( \mathbf{r}\right)  \notag \\
&&-\omega _{0}\int d^{2}\mathbf{r}\left[ \hat{\psi}_{\uparrow }^{\dagger
}\left( \mathbf{r}\right) \hat{\psi}_{\uparrow }\left( \mathbf{r}\right) -%
\hat{\psi}_{\downarrow }^{\dagger }\left( \mathbf{r}\right) \hat{\psi}%
_{\downarrow }\left( \mathbf{r}\right) \right]  \notag \\
&&+\frac{\eta }{\sqrt{N}}\int d^{2}\mathbf{r}\left[ \hat{\psi}_{\downarrow
}^{\dagger }\left( \mathbf{r}\right) \hat{\psi}_{\uparrow }\left( \mathbf{r}%
\right) +\hat{\psi}_{\uparrow }^{\dagger }\left( \mathbf{r}\right) \hat{\psi}%
_{\downarrow }\left( \mathbf{r}\right) \right] \left( \hat{a}+\hat{a}%
^{\dagger }\right)  \notag \\
&&+\lambda \int d^{2}\mathbf{r}\hat{\psi}_{\uparrow }^{\dagger }\left(
\mathbf{r}\right) \hat{\psi}_{\downarrow }^{\dagger }\left( \mathbf{r}%
\right) \hat{\psi}_{\downarrow }\left( \mathbf{r}\right) \hat{\psi}%
_{\uparrow }\left( \mathbf{r}\right) ,  \label{S11}
\end{eqnarray}%
where the factor $1/\sqrt{N}$, with $N$ being the total atom number, is
introduced to ensure that the free energy per fermion is finite in the
thermodynamic limit \cite{YKW73,FTH73}. In the Hamiltonian (\ref{S11}), $%
\omega _{0}=$ $\left( \tilde{\omega}_{\downarrow }-\tilde{\omega}_{\uparrow
}\right) /2$ is the effective resonant frequency between the ground states $%
\left\vert \uparrow \right\rangle $ and $\left\vert \downarrow \right\rangle
$ and can be usually regarded as an effective Zeeman field. For simplicity,
we take $\omega _{0}>0$ in the following discussions. $\eta =\sqrt{N}%
g_{1}\Omega _{1}^{\ast }/\left( 2\Delta _{1}\right) $ $=\sqrt{N}g_{2}\Omega
_{2}^{\ast }/\left( 2\Delta _{2}\right) $ is the effective fermion-photon
coupling strength. $\omega =N\zeta +\tilde{\omega}$ is the atom-number
dependent cavity frequency, where $\zeta =|g_{1}|^{2}/\Delta
_{1}=|g_{2}|^{2}/\Delta _{2}$. It should be emphasized that the choice of
parameters in Eq.~(\ref{S10}) has been used experimentally \cite%
{KB10,KB11,MPB14}.

The Hamiltonian (\ref{S11}) is our required Hamiltonian that governs two
fundamental interactions, including the fermion-photon and fermion-fermion
two-body interactions, and thus, is called a generalized Fermi-Dicke model.
This Hamiltonian has a distinct advantage that all parameters can be
controlled independently. For example, $\omega _{0}$ and $\omega $ can be
tuned by controlling the frequencies of the driving and transverse pumping
lasers, and $\eta $ can be determined by the Rabi frequencies of the
transverse pumping lasers. Besides, $\lambda $ can be tuned by varying the $%
s $-wave scattering length $a_{s}$ through the Feshbach resonant technique
\cite{CC10}. See more detailed discussions in Sec.~\ref{Parameter estimation}%
.\newline

\section{Ground-state properties}

\label{Ground-state}

In order to investigate the ground-state properties of the Hamiltonian (\ref%
{S11}), we first expand operators of the fermionic fields in terms of the
plane waves, i.e.,
\begin{equation}
\hat{\psi}_{\sigma }\left( \mathbf{r}\right) =\frac{1}{\sqrt{S}}\sum_{%
\mathbf{k,}\sigma =\uparrow ,\downarrow }\hat{C}_{\mathbf{k},\sigma }e^{i%
\mathbf{k\cdot r}},  \label{PWF}
\end{equation}%
where $\hat{C}_{\mathbf{k},\sigma }$ are the annihilation operators of
fermions in the momentum space and $S$ is the gas area (hereafter $S=1$).
After a straightforward calculation, we obtain
\begin{widetext}
\begin{eqnarray}
\hat{H} &=&\omega \hat{a}^{\dag }\hat{a}+\sum_{\mathbf{k}}\xi _{\mathbf{k}}%
\hat{C}_{\mathbf{k},\sigma }^{\dagger }\hat{C}_{\mathbf{k},\sigma }+\omega
_{0}\sum_{\mathbf{k}}\left( \hat{C}_{\mathbf{k},\uparrow }^{\dagger }\hat{C}%
_{\mathbf{k},\uparrow }-\hat{C}_{\mathbf{k},\downarrow }^{\dagger }\hat{C}_{%
\mathbf{k},\downarrow }\right)  \notag \\
&&+\lambda \sum_{\mathbf{k}}\hat{C}_{\mathbf{k},\uparrow }^{\dagger }\hat{C}%
_{-\mathbf{k},\downarrow }^{\dagger }\hat{C}_{-\mathbf{k},\downarrow }\hat{C}%
_{\mathbf{k},\uparrow }+\frac{\eta }{\sqrt{n}}\sum_{\mathbf{k}}\left( \hat{C}%
_{\mathbf{k},\uparrow }^{\dagger }\hat{C}_{\mathbf{k},\downarrow }+\hat{C}_{%
\mathbf{k},\downarrow }^{\dagger }\hat{C}_{\mathbf{k},\uparrow }\right)
\left( \hat{a}+\hat{a}^{\dagger }\right),  \label{MSH}
\end{eqnarray}%
\end{widetext}where $\xi _{\mathbf{k}}=\epsilon _{\mathbf{k}}-\mu $, $%
\epsilon _{\mathbf{k}}=\mathbf{k}^{2}/2M$ is the kinetic energy, $%
n=K_{F}^{2}/\left( 2\pi \right) =E_{F}M/\pi $ is the density of fermions in
2D, and $E_{F}=K_{F}^{2}/\left( 2M\right) $ is the Fermi energy. The
Hamiltonian (\ref{MSH}) describes the interaction between two-component
ultracold fermions and a high-finesse optical cavity in the momentum space.

For the attractive fermion-fermion two-body interaction, the Cooper pairing
with the opposite momentum and different spin is formed near to the Fermi
surface \cite{LNC56}. In the mean-field approximation, the corresponding SF
order parameter called the gap is assumed as \cite{MR89,MR90}%
\begin{equation}
\Delta =\lambda \sum_{\mathbf{k}}\left\langle \hat{C}_{-\mathbf{k}%
,\downarrow }\hat{C}_{\mathbf{k},\uparrow }\right\rangle .  \label{SFP}
\end{equation}%
In such case, the two-body interacting Hamiltonian becomes
\begin{equation}
\hat{H}_{\text{INT}}=\Delta \sum_{\mathbf{k}}\left( \hat{C}_{-\mathbf{k}%
,\downarrow }\hat{C}_{\mathbf{k},\uparrow }+\hat{C}_{\mathbf{k},\uparrow
}^{\dagger }\hat{C}_{-\mathbf{k},\downarrow }^{\dagger }\right) -\frac{%
\Delta ^{2}}{\lambda }.  \label{HINT}
\end{equation}%
For simplicity, the mean-field gap is here assumed to be real, i.e., $\Delta
=\Delta ^{\ast }$.

In addition, our considered system usually exists the cavity decay with rate
$\kappa $, and thus, we should introduce the Hesienberg-Langevin equation
for the cavity field operator $\hat{a}$ \cite{GWF88,MOS97},%
\begin{equation}
i\frac{\partial \hat{a}}{\partial t}=\left[ \hat{a},\hat{H}\right] -i\kappa
\hat{a}+\hat{\gamma}_{\text{in}}\left( t\right) ,  \label{HE}
\end{equation}%
where $\hat{\gamma}_{\text{in}}\left( t\right) $ is the quantum noise
operator and satisfies the following conditions: $\left\langle \hat{\gamma}_{%
\text{in}}^{\dagger }\left( t\right) \hat{\gamma}_{\text{in}}\left(
t^{\prime }\right) \right\rangle =2\kappa \delta \left( t-t^{\prime }\right)
$ and $\left\langle \hat{\gamma}_{\text{in}}\left( t\right) \hat{\gamma}_{%
\text{in}}\left( t^{\prime }\right) \right\rangle =0$. In general, the
fluctuation of quantum noise varies faster than $1/\kappa $ on the time
scale \cite{JL08}. When the time scale of the atom dynamics in the motional
degree of freedom is larger than $1/\kappa $, the cavity field can reach a
steady state \cite{DN08,KB10}, which is responsible for obtaining the
ground-state phase diagrams. In terms of Eqs.~(\ref{MSH}) and (\ref{HE}),
the steady-state solution of $\hat{a}$ is given by
\begin{equation}
\alpha =\left\langle \hat{a}\right\rangle =\frac{\eta \sum_{k}\left\langle
\hat{C}_{\mathbf{k},\uparrow }^{\dagger }\hat{C}_{\mathbf{k},\downarrow }+%
\hat{C}_{\mathbf{k},\downarrow }^{\dagger }\hat{C}_{\mathbf{k},\uparrow
}\right\rangle }{\sqrt{n}\left( -\omega +i\kappa \right) }.  \label{SRO}
\end{equation}%
Notice that when considering $\left\langle \hat{a}\right\rangle $, the noise
term can be neglected. In experiments \cite{KB10,KB11}, the mean-photon
number $\left\langle \hat{a}^{\dagger }\hat{a}\right\rangle =\left\vert
\alpha \right\vert ^{2}$ governs the SR properties and is thus called the SR
order parameter.

Based on above discussions and in the basis of Nambu spinor $\hat{\Psi}_{%
\mathbf{k}}=\left( \hat{C}_{\mathbf{k},\uparrow },\hat{C}_{\mathbf{k}%
,\downarrow },\hat{C}_{-\mathbf{k},\downarrow }^{\dagger },-\hat{C}_{-%
\mathbf{k},\uparrow }^{\dagger }\right) ^{T}$, where $T$ stands for the
transposition of a matrix, the Hamiltonian (\ref{MSH}) turns into%
\begin{equation}
\hat{H}=\frac{1}{2}\sum_{\mathbf{k}}\hat{\Psi}_{\mathbf{k}}^{\dagger }M_{%
\mathbf{k}}\hat{\Psi}_{\mathbf{k}}+\sum_{\mathbf{k}}\xi _{\mathbf{k}}-\frac{%
\Delta ^{2}}{\lambda }+\omega \left\vert \alpha \right\vert ^{2},  \label{MH}
\end{equation}%
where the photon-number dependent BdG matrix is given by
\begin{equation}
M_{\mathbf{k}}=\left(
\begin{array}{cccc}
\xi _{\mathbf{k}}-\omega _{0} & \bar{\eta} & \Delta & 0 \\
\bar{\eta} & \xi _{\mathbf{k}}+\omega _{0} & 0 & \Delta \\
\Delta & 0 & -\xi _{\mathbf{k}}-\omega _{0} & \bar{\eta} \\
0 & \Delta & \bar{\eta} & -\xi _{\mathbf{k}}+\omega _{0}%
\end{array}%
\right) ,  \label{BDG}
\end{equation}%
with $\bar{\eta}=\eta \left( \alpha +\alpha ^{\ast }\right) /\sqrt{n}$.\emph{%
\ }The BdG matrix (\ref{BDG}) is also written as
\begin{equation}
M_{\mathbf{k}}=\left(
\begin{array}{cc}
H_{0} & \Delta \mathbbm{1} \\
\Delta \mathbbm{1} & -\sigma _{y}H_{0}\sigma _{y}%
\end{array}%
\right) ,  \label{SH}
\end{equation}%
where $H_{0}=\xi _{\mathbf{k}}+\omega _{0}\sigma _{z}+\bar{\eta}\sigma _{x}$%
, $\sigma _{x}$ and $\sigma _{y}$ are the Pauli matrices, and $\mathbbm{1}$
is the $2\times 2$ unit matrix. The property of the BdG matrix (\ref{SH})
implies that the Hamiltonian (\ref{MH}) has the particle-hole symmetry.

By diagonalizing the BdG matrix $M_{\mathbf{k}}$, we obtain the following
dispersion relations of the Bogoliubov quasiparticles:
\begin{equation}
E_{\mathbf{k},\pm }^{\upsilon }=\upsilon \left( \sqrt{\xi _{\mathbf{k}%
}^{2}+\Delta ^{2}}\pm \sqrt{\bar{\eta}^{2}+\omega _{0}^{2}}\right) ,
\label{QEE}
\end{equation}%
where $\upsilon =\pm 1$ correspond to the particle and hole branches of the
excitation spectra, and $\bar{\eta}^{2}=4\omega ^{2}\eta ^{2}\left\vert
\alpha \right\vert ^{2}/\left[ n\left( \omega ^{2}+\kappa ^{2}\right) \right]
$. For each branch, there are two different excitations, due to the
coexistence of the fermion-photon and fermion-fermion two-body interactions.
In terms of Eq.~(\ref{QEE}), the Hamiltonian (\ref{MH}) is rewritten as
\begin{eqnarray}
\hat{H} &=&\sum_{\mathbf{k}}\left( E_{\mathbf{k},+}\hat{\alpha}_{\mathbf{k}%
,+}^{\dagger }\hat{\alpha}_{\mathbf{k},+}+E_{\mathbf{k},-}\hat{\alpha}_{%
\mathbf{k},-}^{\dagger }\hat{\alpha}_{\mathbf{k},-}\right)  \notag \\
&&+\sum_{\mathbf{k}}\left( \xi _{\mathbf{k}}-\sqrt{\xi _{\mathbf{k}%
}^{2}+\Delta ^{2}}\right) -\frac{\Delta ^{2}}{\lambda }+\omega \left\vert
\alpha \right\vert ^{2},  \label{DH}
\end{eqnarray}%
where $\hat{\alpha}_{\mathbf{k},\pm }$ are the operators of the Bogoliubov
quasiparticles and satisfy the anticommutation relations $\left\{ \hat{\alpha%
}_{\mathbf{k},\pm }^{\dagger },\hat{\alpha}_{\mathbf{k},\pm }\right\}
=\delta _{ll^{\prime }}$ ($l=l^{\prime }=\pm $).

If $E_{\mathbf{k},+}$ and $E_{\mathbf{k},-}$ in Eq.~(\ref{QEE}) are both
positive, the first two terms in the Hamiltonian (\ref{DH}) reflect the
excitation energies and its rest term is called the ground-state energy. In
fact, $E_{\mathbf{k},\pm }$ are positive definite only when $\sqrt{\xi _{%
\mathbf{k}}^{2}+\Delta ^{2}}>\sqrt{\bar{\eta}^{2}+\omega _{0}^{2}}$. In
order to correctly write down the Hamiltonian (\ref{DH}) as a sum of the
excitation energies and the ground-state energy, we should introduce the
Heaviside step function, which is defined as $\Theta \left( x\right) =1$ for
$x\geqslant 0$ and $\Theta \left( x\right) =0$ for $x<0$. This Heaviside
step function can help us separate the sum over momenta in the different
regions \cite{DES07}: $E_{\mathbf{k},\pm }>0$ and $E_{\mathbf{k},\pm }<0$.
By means of $\Theta \left( x\right) $, the Hamiltonian (\ref{DH}) becomes
\begin{eqnarray}
\hat{H} &=&\sum_{\mathbf{k,\pm }}E_{\mathbf{k},\pm }\Theta \left( E_{\mathbf{%
k},\pm }\right) \hat{\alpha}_{\mathbf{k},\pm }^{\dagger }\hat{\alpha}_{%
\mathbf{k},\pm }  \notag \\
&&-\sum_{\mathbf{k,\pm }}E_{\mathbf{k},\pm }\Theta \left( -E_{\mathbf{k},\pm
}\right) \hat{\alpha}_{\mathbf{k},\pm }\hat{\alpha}_{\mathbf{k},\pm
}^{\dagger }+E_{\text{G}},  \label{RH}
\end{eqnarray}%
with
\begin{eqnarray}
E_{\text{G}} &=&\sum_{\mathbf{k}}\left[ E_{\mathbf{k},+}\Theta \left( -E_{%
\mathbf{k},+}\right) +E_{\mathbf{k},-}\Theta \left( -E_{\mathbf{k},-}\right) %
\right]  \notag \\
&&+\sum_{\mathbf{k}}\left( \xi _{\mathbf{k}}-\sqrt{\xi _{\mathbf{k}%
}^{2}+\Delta ^{2}}\right) -\frac{\Delta ^{2}}{\lambda }+\omega \left\vert
\alpha \right\vert ^{2}.  \label{GSE}
\end{eqnarray}%
In terms of Eq.~(\ref{QEE}), it is easy to see that $E_{\mathbf{k},+}$ is
always positive, i.e., $\Theta \left( E_{\mathbf{k},+}\right) =1$.
Therefore, the ground-state energy in Eq.~(\ref{GSE}) is simplified as a
simple form%
\begin{eqnarray}
E_{\text{G}} &=&\sum_{\mathbf{k}}E_{\mathbf{k},-}\Theta \left( -E_{\mathbf{k}%
,-}\right)  \notag \\
&&+\sum_{\mathbf{k}}\left( \xi _{\mathbf{k}}-\sqrt{\xi _{\mathbf{k}%
}^{2}+\Delta ^{2}}\right) -\frac{\Delta ^{2}}{\lambda }+\omega \left\vert
\alpha \right\vert ^{2}.  \label{SGE}
\end{eqnarray}

According to the ground-state energy in Eq.~(\ref{SGE}), three parameters $%
\Delta $, $\mu $, and $\left\vert \alpha \right\vert $ can be derived from
the mean-field gap equation $\partial E_{\text{G}}/\partial \Delta =0$, the
particle number equation $\partial E_{\text{G}}/\partial \mu =-n$, and the
SR equation $\partial E_{\text{G}}/\partial \left( \left\vert \alpha
\right\vert \right) =0$. If using the relation $\Theta ^{\prime }(x)=\delta
\left( x\right) $, where $\delta \left( x\right) $ is the Dirac delta
function, the above three equations are given respectively by
\begin{eqnarray}
\Delta \left[ \sum_{\mathbf{k}}\frac{f(-E_{\mathbf{k},-})}{\sqrt{\xi _{%
\mathbf{k}}^{2}+\Delta ^{2}}}-\sum_{\mathbf{k}}\frac{1}{\sqrt{\xi _{\mathbf{k%
}}^{2}+\Delta ^{2}}}-\frac{2}{\lambda }\right] &=&0,  \label{GE} \\
\sum_{\mathbf{k}}\frac{\xi _{\mathbf{k}}f(-E_{\mathbf{k},-})}{\sqrt{\xi _{%
\mathbf{k}}^{2}+\Delta ^{2}}}+\sum_{\mathbf{k}}\left( 1-\frac{\xi _{\mathbf{k%
}}}{\sqrt{\xi _{\mathbf{k}}^{2}+\Delta ^{2}}}\right) &=&n,  \label{ANE} \\
\left\vert \alpha \right\vert \left[ \sum_{\mathbf{k}}-2\omega \eta
^{2}f(-E_{\mathbf{k},-})+n\bar{\chi}\left( \omega ^{2}+\kappa ^{2}\right) %
\right] &=&0,  \label{PDE}
\end{eqnarray}%
where $f(-E_{\mathbf{k},-})=\Theta \left( -E_{\mathbf{k},-}\right) -E_{%
\mathbf{k},-}\delta \left( -E_{\mathbf{k},-}\right) $ and $\bar{\chi}=\sqrt{%
\bar{\eta}^{2}+\omega _{0}^{2}}$. Notice that when $k\rightarrow \mathbf{%
\infty }$, Eq.~(\ref{GE}) diverges. In order to eliminate this ultraviolet
divergence, $\lambda $ should be renormalized as \cite{MR89,MR90}
\begin{equation}
\frac{1}{\lambda }=-\sum_{\mathbf{k}}\frac{1}{2\epsilon _{\mathbf{k}}+E_{b}},
\label{ISR}
\end{equation}%
where $E_{b}>0$ is the two-body binding energy in 2D. In the following
discussions, we self-consistently solve the coupled equations (\ref{GE})-(%
\ref{PDE}) at a fixed atom density $n$ to obtain three parameters $\Delta $,
$\mu $, and $\left\vert \alpha \right\vert $. Equations (\ref{GE}) and (\ref%
{PDE}) show that there always exists different solutions about $\Delta $ and
$\left\vert \alpha \right\vert $. In fact, we must consider the stability of
the system to find the proper solutions, and then predict rich quantum phase
and phase diagrams. For simplicity, we take $E_{F}$ as the unit of energy.%
\newline

\section{Phase diagrams for $E_{b}=0$}

\label{Phase I}

When $E_{b}=0$, the system has no fermion-fermion two-body interaction. To
better understand the relevant behavior, we first consider the case of $\eta
=0$, i.e., the free Fermi gas. In this case, the scaled ground-state energy
defined as $\bar{E}_{\text{G}}=E_{\text{G}}/n$ (i.e., the ground-state
energy per fermion) is given by
\begin{equation}
\bar{E}_{\text{G}}\!=\!-\frac{1}{4E_{F}}\!\left[ \left( \mu \!+\!\omega
_{0}\right) ^{2}\Theta \left( \mu \!+\!\omega _{0}\right) \!+\!\left( \mu
\!-\!\omega _{0}\right) ^{2}\Theta \left( \mu \!-\!\omega _{0}\right) \right]
.  \label{GN}
\end{equation}%
We always assume $\Theta \left( \mu +\omega _{0}\right) \neq 0$, which
implies $\mu +\omega _{0}\;>\;0$. If $\mu +\omega _{0}\;<\;0$, $\bar{E}_{%
\text{G}}=0$, in which the system has no definite physical meaning since
under such condition no real fermions can be found \cite{DES15}. When $%
\omega _{0}>\mu $, $\Theta \left( \mu -\omega _{0}\right) =0$ and the scaled
ground-state energy in Eq.~(\ref{GN}) becomes%
\begin{equation}
\bar{E}_{\text{G}}=-\frac{\left( \omega _{0}+\mu \right) ^{2}}{4E_{F}}.
\label{GN1}
\end{equation}%
In order to describe the effects induced by the effective Zeeman field, we
should introduce the scaled polarization \cite{DES06}
\begin{equation}
\bar{m}=\frac{n_{\uparrow }-n_{\downarrow }}{n}=-\frac{\partial \bar{E}_{%
\text{G}}}{\partial \omega _{0}}.  \label{MPE}
\end{equation}%
Based on Eqs.~(\ref{ANE}),~(\ref{GN1}), and~(\ref{MPE}), we obtain
\begin{equation}
\mu =2E_{F}-\omega _{0}\text{, \ \ \ \ \ }\bar{m}=1.  \label{NIG}
\end{equation}%
Equation (\ref{NIG}) shows that the Fermi gas is fully polarized and the
system only has a Fermi surface with $\mu _{\uparrow }=2E_{F}$. The
corresponding normal phase is called the N-I phase. When $\omega
_{0}\;<\;\mu $, $\Theta \left( \mu -\omega _{0}\right) =1$ and the scaled
ground-state energy in Eq.~(\ref{GN}) becomes%
\begin{equation}
\bar{E}_{\text{G}}=-\frac{\left( \mu ^{2}+\omega _{0}^{2}\right) }{2E_{F}}.
\label{GN2}
\end{equation}%
We further obtain
\begin{equation}
\text{\ \ }\mu =E_{F}\text{, \ \ \ \ \ }\bar{m}=\frac{\omega _{0}}{E_{F}}.
\label{NTG}
\end{equation}%
It is quite different from the N-I phase that in this case the Fermi gas is
partially polarized and the system has two Fermi surfaces defined
respectively as $\mu _{\uparrow }=E_{F}+\omega _{0}$\ and $\mu _{\downarrow
}=E_{F}-\omega _{0}$. The corresponding phase is called the N-II phase. From
above discussions, it can be seen that when varying $\omega _{0}$, the
system undergoes a first-order phase transition from the N-I phase to the
N-II phase at the critical point $\omega _{0}^{c}=E_{F}$ \cite{DES07}.

For a weak $\eta $, the noninteracting terms in the Hamiltonian (\ref{MSH}),
$\omega \hat{a}^{\dagger }\hat{a}+\sum_{\mathbf{k}}\xi _{\mathbf{k}}\hat{C}_{%
\mathbf{k,\sigma }}^{\dagger }\hat{C}_{\mathbf{k,\sigma }}+\omega _{0}\sum_{%
\mathbf{k}}\left( \hat{C}_{\mathbf{k,\uparrow }}^{\dagger }\hat{C}_{\mathbf{%
k,\uparrow }}-\hat{C}_{\mathbf{k,\downarrow }}^{\dagger }\hat{C}_{\mathbf{%
k,\downarrow }}\right) $, play a dominate role in the systematic dynamics.
In this case, no fermion-photon interaction occurs and $\left\vert \alpha
\right\vert ^{2}=0$ for the ground state, i.e., the system still remains the
fundamental properties of the N-I or N-II phases.

If $\eta $\ becomes stronger, the fermion-photon interacting term, $\left(
\eta /\sqrt{n}\right) \sum_{\mathbf{k}}\left( \hat{C}_{\mathbf{k,\uparrow }%
}^{\dagger }\hat{C}_{\mathbf{k,\downarrow }}+\hat{C}_{\mathbf{k,\downarrow }%
}^{\dagger }\hat{C}_{\mathbf{k,\uparrow }}\right) \left( \hat{a}+\hat{a}%
^{\dagger }\right) $, dominates and the system acquires the macroscopic
collective excitation with $\left\vert \alpha \right\vert ^{2}\neq 0$. This
implies that when increasing $\eta $, the SR transition can be expected to
occur. In terms of Eq.~(\ref{SGE}), the corresponding scaled ground-state
energy is obtained by%
\begin{equation}
\bar{E}_{\text{G}}=-\frac{1}{4E_{F}}\left[ \mu _{+}^{2}+\mu _{-}^{2}\Theta
\left( \mu _{-}\right) \right] +\omega \left\vert \bar{\alpha}\right\vert
^{2},  \label{SGE1}
\end{equation}%
where $\mu _{\pm }=\mu \pm \bar{\chi}$ and $\left\vert \bar{\alpha}%
\right\vert ^{2}=\left\vert \alpha \right\vert ^{2}/n$ is the scaled
mean-photon number. Since the Heaviside step function in Eq.~(\ref{SGE1})
depends crucially on $\mu _{-}$, the following discussion of the
ground-state properties should be divided into two specific cases: $\mu
_{-}<0$ and $\mu _{-}\geqslant 0$.

\subsection{$\protect\mu _{-}<0$}

When
\begin{equation}
\mu _{-}=\mu -\bar{\chi}<0,  \label{TH}
\end{equation}%
$\Theta \left( \mu _{-}\right) =0$ and the scaled ground-state energy in
Eq.~(\ref{SGE1}) becomes
\begin{equation}
\bar{E}_{\text{G}}=-\frac{1}{4E_{F}}\mu _{+}^{2}+\omega \left\vert \bar{%
\alpha}\right\vert ^{2}.  \label{SGE2}
\end{equation}%
In addition, Eqs.~(\ref{ANE}), (\ref{PDE}), and (\ref{MPE}) turn into
\begin{eqnarray}
\mu +\sqrt{\bar{\eta}^{2}+\omega _{0}^{2}} &=&2E_{F},  \label{Cp} \\
\left\vert \bar{\alpha}\right\vert \left[ 2\omega -\frac{2\omega ^{2}\eta
^{2}\left( \mu +\sqrt{\bar{\eta}^{2}+\omega _{0}^{2}}\right) }{E_{F}\left(
\omega ^{2}+\kappa ^{2}\right) \sqrt{\bar{\eta}^{2}+\omega _{0}^{2}}}\right]
&=&0,  \label{CFP} \\
\frac{\omega _{0}\left( \mu +\sqrt{\bar{\eta}^{2}+\omega _{0}^{2}}\right) }{%
2E_{F}\sqrt{\bar{\eta}^{2}+\omega _{0}^{2}}} &=&\bar{m}.  \label{PL}
\end{eqnarray}%
By further solving Eqs.~(\ref{Cp})-(\ref{PL}), we obtain%
\begin{equation}
\text{ }\mu =2E_{F}-\omega _{0}\text{, \ \ \ }\left\vert \bar{\alpha}%
\right\vert =0\text{, \ \ \ }\bar{m}=1,  \label{Sc1}
\end{equation}%
or%
\begin{equation}
\left\{
\begin{array}{l}
\mu =2\left( E_{F}-\frac{\omega \eta ^{2}}{\omega ^{2}+\kappa ^{2}}\right) ,
\\
\left\vert \bar{\alpha}\right\vert =\sqrt{\frac{\eta ^{2}}{\omega
^{2}+\kappa ^{2}}-\frac{\omega _{0}^{2}\left( \omega ^{2}+\kappa ^{2}\right)
}{4\omega ^{2}\eta ^{2}}}, \\
\text{ }\bar{m}=\frac{\omega _{0}\left( \omega ^{2}+\kappa ^{2}\right) }{%
2\omega \eta ^{2}}.%
\end{array}%
\right.  \label{Cs}
\end{equation}

\begin{figure}[tbh]
\centering
\includegraphics[width=5cm]{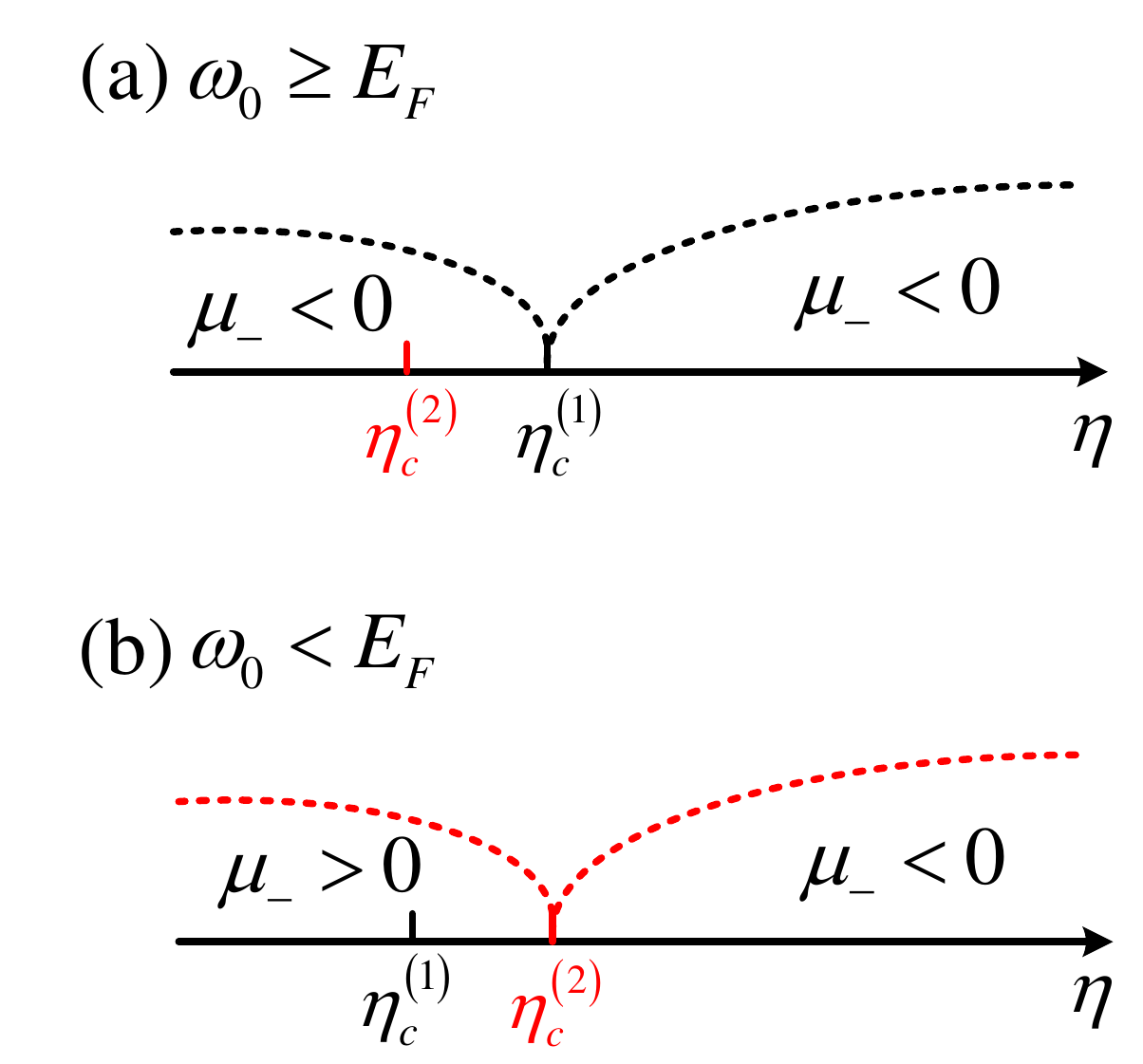}\newline
\caption{The comparison of the critical points $\protect\eta _{c}^{(1)}$ and
$\protect\eta _{c}^{(2)}$ for (a) $\protect\omega _{0}\geqslant E_{F}$ and
(b) $\protect\omega _{0}<E_{F}$. When $\protect\omega _{0}\geqslant E_{F}$, $%
\protect\eta _{c}^{(1)}>\protect\eta _{c}^{(2)}$, and $\protect\mu _{-}<0$
for both $0<\protect\eta <\protect\eta _{c}^{(1)}$ and $\protect\eta >%
\protect\eta _{c}^{(1)}$. When $\protect\omega _{0}<E_{F}$, $\protect\eta %
_{c}^{(1)}<\protect\eta _{c}^{(2)}$, and $\protect\mu _{-}<0$\ for $\protect%
\eta >\protect\eta _{c}^{(2)}$ and $\protect\mu _{-}>0$ for $0<\protect\eta <%
\protect\eta _{c}^{(2)}$.}
\label{figCP}
\end{figure}
In order to find the stable ground state, we should introduce the condition
governed by $\partial ^{2}\bar{E}_{\text{G}}/\partial \left( \left\vert \bar{%
\alpha}\right\vert \right) ^{2}>0$. In terms of this stable condition, we
find immediately that $\left\vert \bar{\alpha}\right\vert =0$ for $\eta
<\eta _{c}^{(1)}$, where
\begin{equation}
\eta _{c}^{(1)}=\sqrt{\frac{\omega _{0}\left( \omega ^{2}+\kappa ^{2}\right)
}{2\omega }},  \label{CC}
\end{equation}%
and $\left\vert \bar{\alpha}\right\vert =\sqrt{\eta ^{2}/\left( \omega
^{2}+\kappa ^{2}\right) -\omega _{0}^{2}\left( \omega ^{2}+\kappa
^{2}\right) /\left( 4\omega ^{2}\eta ^{2}\right) }$ [i.e., Eq.~(\ref{Cs})]
for $\eta >\eta _{c}^{(1)}$. At the same time, we must notice the
restrictive condition $\mu _{-}=\mu -\bar{\chi}<0$ in Eq.~(\ref{TH}), which
also induces a critical point
\begin{equation}
\eta _{c}^{(2)}=\sqrt{\frac{E_{F}\left( \omega ^{2}+\kappa ^{2}\right) }{%
2\omega }}.  \label{CC2}
\end{equation}%
When $\eta >\eta _{c}^{(2)}$, $\mu _{-}<0$.

From the critical points in Eqs.~(\ref{CC}) and (\ref{CC2}), we find that
when $\omega _{0}\geqslant E_{F}$, $\eta _{c}^{(1)}>\eta _{c}^{(2)}$, and
Eqs.~(\ref{Sc1})-(\ref{Cs}) satisfy the condition $\mu _{-}<0$ for both $%
0<\eta <\eta _{c}^{(1)}$ and $\eta >\eta _{c}^{(1)}$ [see Fig.~\ref{figCP}%
(a)]. However, when $\omega _{0}<E_{F}$, $\eta _{c}^{(1)}<\eta _{c}^{(2)}$,
and only Eq.~(\ref{Cs}) satisfy the condition $\mu _{-}<0$ for $\eta >\eta
_{c}^{(2)}$ [see Fig.~\ref{figCP}(b)]. In other words, when $0<\eta <\eta
_{c}^{(2)}$, $\mu _{-}\geqslant 0$, in which we should introduce a new
scaled ground-state energy, as will be shown in the next subsection.

As a consequence, in the case of $\omega _{0}\geqslant E_{F}$, the scaled
ground-state energy is written as%
\begin{equation}
\bar{E}_{\text{G}}=\left\{
\begin{array}{lcl}
{-E_{F}} &  & {0<\eta <\eta _{c}^{(1)}} \\
{-E_{F}+\frac{\omega \eta ^{2}}{\omega ^{2}+\kappa ^{2}}-\frac{\omega
_{0}^{2}\left( \omega ^{2}+\kappa ^{2}\right) }{4\omega \eta ^{2}}} &  & {%
\eta >\eta _{c}^{(1)}}%
\end{array}%
\right. ,  \label{LAG}
\end{equation}%
and three parameters are given respectively by%
\begin{eqnarray}
\mu &=&\left\{
\begin{array}{lcl}
{2E_{F}-\omega _{0}} &  & {0<\eta <\eta _{c}^{(1)}} \\
{2\left( E_{F}-\frac{\omega \eta ^{2}}{\omega ^{2}+\kappa ^{2}}\right) } &
& {\eta >\eta _{c}^{(1)}}%
\end{array}%
\right.,  \label{LAC} \\
\left\vert \bar{\alpha}\right\vert &=&\left\{
\begin{array}{lcl}
0 &  & {0<\eta <\eta _{c}^{(1)}} \\
\sqrt{\frac{\eta ^{2}}{\omega ^{2}+\kappa ^{2}}-\frac{\omega _{0}^{2}\left(
\omega ^{2}+\kappa ^{2}\right) }{4\omega ^{2}\eta ^{2}}} &  & \eta >\eta
_{c}^{(1)}%
\end{array}%
\right.,  \label{LAA} \\
\bar{m} &=&\left\{
\begin{array}{lcl}
1 &  & {0<\eta <\eta _{c}^{(1)}} \\
\frac{\omega _{0}\left( \omega ^{2}+\kappa ^{2}\right) }{2\omega \eta ^{2}}
&  & {\eta >\eta _{c}^{(1)}}%
\end{array}%
\right..  \label{LAm}
\end{eqnarray}

\begin{figure}[tbh]
\centering
\includegraphics[width=8cm]{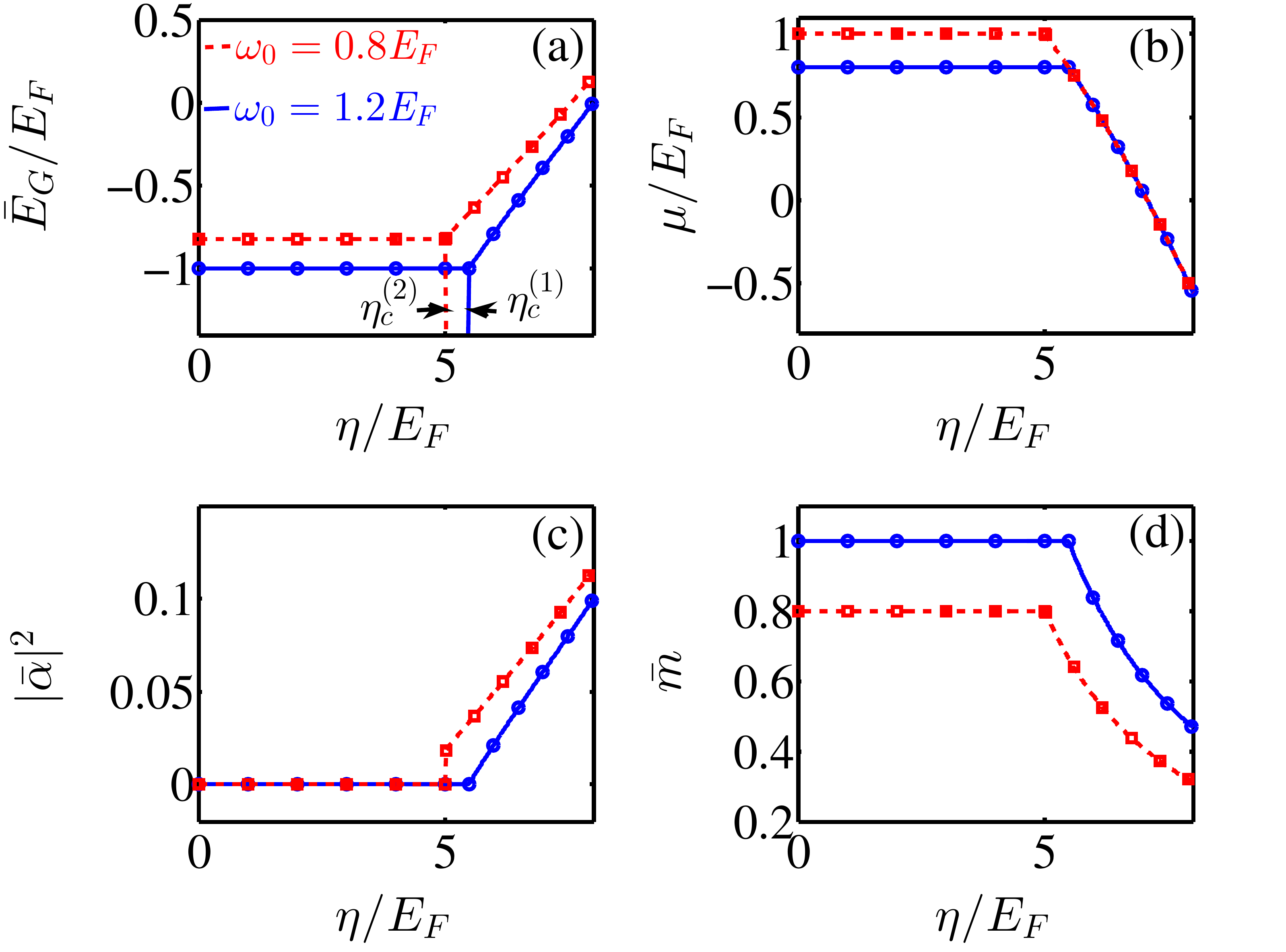}\newline
\caption{(a) The scaled ground-state energy $\bar{E}_{\text{G}}/E_{F}$, (b)
the chemical potential $\protect\mu /E_{F}$, (c) the scaled mean-photon
number $\left\vert \bar{\protect\alpha}\right\vert ^{2}$, and (d) the scaled
polarization $\bar{m}$ as functions of the effective atom-photon coupling
strength $\protect\eta /E_{F}$. The atom-number dependent cavity frequency
and the cavity decay rate are given by $\protect\omega =10E_{F}$ and $%
\protect\kappa =20E_{F}$, respectively. When the effective resonant
frequency is chosen as $\;\protect\omega _{0}=1.2E_{F}$, $\protect\eta %
_{c}^{(1)}=5.48E_{F}$. When $\;\protect\omega _{0}=0.8E_{F}$, $\protect\eta %
_{c}^{(2)}=$ $5E_{F}$. In these figures, the blue-solid and red-dashed lines
represent the analytical results, while the open symbols correspond to
numerical simulations.}
\label{figMP}
\end{figure}
The analytical results in Eqs.~(\ref{LAG})-(\ref{LAm}) show two typical
properties. The first is that the first-order derivative of $\bar{E}_{\text{G%
}}$ with respect to ${\eta }$ is continuous but its second order is
discontinuous, which means that a second-order phase transition from the N-I
phase to the SR phase occurs at the critical point $\eta _{c}^{(1)}$. This
property is similar to the case of ultracold Bose atoms \cite{YKW73,FTH73}.
The other is that in the N-I phase no fermion-photon interaction occurs,
whereas in the SR phase both the fermions and photons acquire the
macroscopic collective excitations. Besides, $\bar{m}$ is inversely
proportional to $\eta ^{2}$ and $\mu $ is decreased with respect to $\eta
^{2}$. In Figs.~\ref{figMP}(a)-\ref{figMP}(d), we plot $\bar{E}_{\text{G}}$,
$\mu $, $\left\vert \bar{\alpha}\right\vert ^{2}$, and $\bar{m}$ as
functions of $\eta $. These figures show that our analytical results agree
well with direct numerical simulations. Moreover, the above two typical
properties are recovered naturally.

\subsection{$\protect\mu _{-}\geqslant 0$}

When
\begin{equation}
\mu _{-}=\mu -\bar{\chi}\geqslant 0,  \label{TH2}
\end{equation}%
$\Theta \left( \mu _{-}\right) =1$ and the scaled ground-state energy in
Eq.~(\ref{SGE1}) becomes
\begin{equation}
\bar{E}_{\text{G}}=-\frac{1}{4E_{F}}\left( \mu _{+}^{2}+\mu _{-}^{2}\right)
+\omega \left\vert \bar{\alpha}\right\vert ^{2}.  \label{SGE3}
\end{equation}%
According to the discussions in the subsection A of this section and
considering the stable condition governed by $\partial ^{2}\bar{E}_{\text{G}%
}/\partial \left( \left\vert \bar{\alpha}\right\vert \right) ^{2}>0$, we
obtain%
\begin{equation}
\mu =E_{F}\text{, \ \ \ \ }\left\vert \bar{\alpha}\right\vert =0\text{, \ \
\ \ }\bar{m}=\frac{\omega _{0}}{E_{F}},  \label{MYI}
\end{equation}%
for $0<\eta <\eta _{c}^{(2)}$. Since Eq.~(\ref{MYI}) should satisfy the
condition $\mu _{-}\geqslant 0$ when $0<\eta <\eta _{c}^{(2)}$, we find that
$\omega _{0}<E_{F}$ in this case. When $\eta >\eta _{c}^{(2)}$, $\mu _{-}<0$
[see Fig.~\ref{figCP}(b)], in which we should combine with the previous
discussions in the subsection A of this section.

As a consequence, in the case of $\omega _{0}<E_{F}$, the scaled
ground-state energy is written as%
\begin{equation}
\bar{E}_{\text{G}}=\left\{
\begin{array}{lcl}
{-\frac{E_{F}}{2}-\frac{\omega _{0}^{2}}{2E_{F}}} &  & {0<\eta <\eta
_{c}^{(2)}} \\
{-E_{F}+\frac{\omega \eta ^{2}}{\omega ^{2}+\kappa ^{2}}-\frac{\omega
_{0}^{2}\left( \omega ^{2}+\kappa ^{2}\right) }{4\omega \eta ^{2}}} &  & {%
\eta >\eta _{c}^{(2)}}%
\end{array}%
\right. .  \label{LAM}
\end{equation}%
In addition, when $0<\eta <\eta _{c}^{\left( 2\right) }$, $\mu $, $%
\left\vert \bar{\alpha}\right\vert $, and $\bar{m}$ are governed by Eq.~(\ref%
{MYI}), which means that the system is located at the N-II phase. When $\eta
>\eta _{c}^{(2)}$, $\mu $, $\left\vert \bar{\alpha}\right\vert $, and $\bar{m%
}$\ are governed by Eq.~(\ref{Cs}), which implies that the system is located
at the SR phase.\ The scaled ground-state energy in Eq.~(\ref{LAM}) shows
that the phase transition from the N-II phase to the SR phase is of the
first order. Moreover, $\left\vert \bar{\alpha}\right\vert ^{2}$ exhibits a
sudden change at the critical point $\eta _{c}^{(2)}$, as shown by the
red-dashed line in Fig.~\ref{figMP}(c).

\begin{figure}[!t]
\centering
\includegraphics[width=7cm]{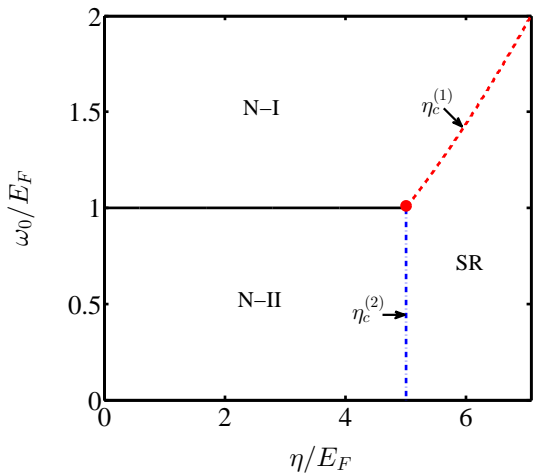}
\caption{Phase diagram as a function of the effective resonant frequency $%
\protect\omega _{0}/E_{F}$ and the effective fermion-photon coupling
strength $\protect\eta /E_{F}$. The atom-number dependent cavity frequency $%
\protect\omega $ and the cavity decay rate $\protect\kappa $ are the same as
those in Fig.~\protect\ref{figMP}. When the effective resonant frequency is
chosen as $\protect\omega _{0}=E_{F}$, $\protect\eta _{c}^{\left( 1\right) }=%
\protect\eta _{c}^{\left( 2\right) }$.}
\label{figPD}
\end{figure}

Interestingly, when $\eta =\eta _{c}^{\left( 2\right) }$, the scaled
ground-state energies of the N-II and SR phases are equal, which means that
these two phases coexist and the corresponding phase is called \textit{the
N-II-SR mixed phase}. In order to fully describe the fundamental properties
of this mixed phase, we should introduce the fractions of the N-II and SR
phases, $x_{0}$ and $1-x_{0}$. Moreover, we further obtain
\begin{eqnarray}
\bar{E}_{\text{G}} &=&\mu -\frac{x_{0}\left( \omega
_{0}^{2}+E_{F}^{2}\right) }{2E_{F}}  \notag \\
&&-\frac{\left( 1-x_{0}\right) }{2}\left( E_{F}+\frac{\omega _{0}^{2}}{E_{F}}%
\right) ,  \label{NRG} \\
\mu &=&E_{F},  \label{NRM} \\
\left\vert \bar{\alpha}\right\vert &=&\sqrt{\frac{E_{F}^{2}-\omega _{0}^{2}}{%
2\omega E_{F}}},  \label{NRA} \\
\bar{m} &=&\frac{x_{0}\omega _{0}}{E_{F}}+\left( 1-x_{0}\right) \frac{\omega
_{0}\left( \omega ^{2}+\kappa ^{2}\right) }{2\omega \eta ^{2}}.  \label{NRP}
\end{eqnarray}%
The detailed derivation of Eqs.~(\ref{NRG})-(\ref{NRP}) is given by the
Appendix~\ref{AppAI}. Notice that since the atom densities are equal in both
the N-II and SR phases, $x_{0}$ and $1-x_{0}$ are the arbitrary values
ranging from $0$ to $1$. In addition, since $\bar{m}\neq 0$ in both the N-II
and SR phases, the nonzero polarization in the N-II-SR\ mixed phase\ is
caused by both the macroscopic collective excitation of the fermions and
photons and the effective Zeeman field. We also find two first-order phase
transitions from the N-II-SR phase to the N-II phase or the SR phase. These
results are quite different from those in the N-I phase and the ultracold
Bose atoms \cite{YKW73,FTH73}.

\subsection{Phase diagram}

In Fig.~\ref{figPD}, we plot the whole phase diagram, including the N-I
phase, the N-II phase, the N-II-SR mixed phase, and the SR phase, for $%
\omega _{0}<E_{F}$ and $\omega _{0}\geqslant E_{F}$. As predicted
previously, the phase transition from the N-I phase to the SR phase is of
the second order, whereas the phase transition from the N-II phase to the SR
phase is of the first order, due to the coexistence of the N-II and SR
phases at the critical line. In addition, this phase diagram has a
tricritical point (the red dot), at which the phase transition changes from
the first order to the second order.\newline

\section{Phase diagrams for $E_{b}\neq0$ and $\protect\eta \neq 0$}

\label{Phase II}

When both $\eta $\ and $E_{b}$\ exist, the properties of the Bogoliubov
quasiparticle states are determined by both $\bar{\chi}$ and $\Delta $, as
shown in the Hamiltonian (\ref{DH}). If $\bar{\chi}>\Delta $, the
quasi-particle states are occupied for $\max \{0,\mu -\sqrt{\bar{\chi}%
^{2}-\Delta ^{2}}\}<\epsilon _{\mathbf{k}}<\mu +\sqrt{\bar{\chi}^{2}-\Delta
^{2}}$\ ($\mu +\sqrt{\bar{\chi}^{2}-\Delta ^{2}}>0$), and thus, the scaled
ground-state energy is obtained by
\begin{widetext}
\begin{eqnarray}
\bar{E}_{\text{G}} &=&\bar{E}_{\text{G}}^{\text{SF}}+\omega \left\vert \bar{%
\alpha}\right\vert ^{2}-\frac{1}{4E_{F}}\left[ 2\bar{\chi}\sqrt{\bar{\chi}%
^{2}-\Delta ^{2}}-\Delta ^{2}\ln \left( \frac{\bar{\chi}+\sqrt{\bar{\chi}%
^{2}-\Delta ^{2}}}{\bar{\chi}-\sqrt{\bar{\chi}^{2}-\Delta ^{2}}}\right) %
\right] \Theta \left( \bar{\mu}_{-}\right) \Theta \left( \bar{\chi}%
^{2}-\Delta ^{2}\right)  \notag \\
&&-\frac{1}{4E_{F}}\left[ \bar{\chi}\sqrt{\bar{\chi}^{2}-\Delta ^{2}}+\mu
\left( 2\bar{\chi}-\sqrt{\mu ^{2}+\Delta ^{2}}\right) -\Delta ^{2}\ln \left(
-\frac{\bar{\chi}+\sqrt{\bar{\chi}^{2}-\Delta ^{2}}}{\mu -\sqrt{\mu
^{2}+\Delta ^{2}}}\right) \right] \Theta \left( -\bar{\mu}_{-}\right) \Theta
\left( \bar{\chi}^{2}-\Delta ^{2}\right) ,  \label{SGEE4}
\end{eqnarray}%
\end{widetext}where $\bar{\mu}_{-}=\mu -\sqrt{\bar{\chi}^{2}-\Delta ^{2}}$
and
\begin{eqnarray}
\bar{E}_{\text{G}}^{\text{SF}} &=&\frac{\Delta ^{2}}{4E_{F}}\ln \left( \frac{%
\sqrt{\mu ^{2}+\Delta ^{2}}-\mu }{E_{b}}\right)  \notag \\
&&-\frac{\mu }{4E_{F}}\left( \sqrt{\mu ^{2}+\Delta ^{2}}+\mu \right) -\frac{%
\Delta ^{2}}{8E_{F}}  \label{SF}
\end{eqnarray}%
is the fully-paired ($\bar{m}=0$) SF energy. If $\bar{\chi}<\Delta $, $\bar{E%
}_{\text{G}}=\bar{E}_{\text{G}}^{\text{SF}}$. It can be seen clearly from
Eq.~(\ref{SGEE4}) that the ground-state properties, including $\Delta $, $%
\mu $, and $\left\vert \bar{\alpha}\right\vert ^{2}$, are governed by both $%
\eta $ and $E_{b}$. When $E_{b}=0$, $\bar{\mu}_{-}=\mu -\bar{\chi}$\ and the
corresponding scaled ground-state energy is the same as Eq.~(\ref{SGE1}).
When $\eta =0$, $\bar{\chi}=\omega _{0}$ and Eq.~(\ref{SGEE4}) reduces to
Eq.~(14) in Ref.~\cite{JD09} and Eq.~(8) in Ref.~\cite{HC12}. When $%
E_{b}\neq 0$ and $\eta \neq 0$, a strong competition between the SF and SR
properties occurs. Consequently, rich quantum phases can be predicted.
Similarly, the Heaviside step function in Eq.~(\ref{SGEE4}) depends
crucially on $\bar{\mu}_{-}$ and $\bar{\chi}^{2}-\Delta ^{2}$, and thus, the
following discussion of the ground-state properties should also be divided
into four specific cases: $\bar{\mu}_{-}<0$\ and $\bar{\chi}^{2}-\Delta
^{2}\geqslant 0$, $\bar{\mu}_{-}<0$\ and $\bar{\chi}^{2}-\Delta ^{2}<0$, $%
\bar{\mu}_{-}\geqslant 0$\ and $\bar{\chi}^{2}-\Delta ^{2}\geqslant 0$, and $%
\bar{\mu}_{-}\geqslant 0$\ and \ $\bar{\chi}^{2}-\Delta ^{2}<0$. Moreover,
we will draw two conclusions for $\omega _{0}<E_{F}$ and $\omega
_{0}\geqslant E_{F}$.

\subsection{$\bar{\protect\mu}_{-}<0$ and $\bar{\protect\chi}^{2}-\Delta
^{2}\geqslant 0$}

When%
\begin{equation}
\bar{\mu}_{-}=\mu -\sqrt{\bar{\chi}^{2}-\Delta ^{2}}<0  \label{TH3}
\end{equation}%
and
\begin{equation}
\bar{\chi}^{2}-\Delta ^{2}\geqslant 0,  \label{TH3A}
\end{equation}%
$\Theta \left( \bar{\mu}_{-}\right) =0$ and $\Theta \left( \bar{\chi}%
^{2}-\Delta ^{2}\right) =1$. Thus, the scaled ground-state energy in Eq.~(%
\ref{SGEE4}) becomes
\begin{eqnarray}
\bar{E}_{\text{G}} &=&\bar{E}_{\text{G}}^{\text{SF}}+\omega \left\vert \bar{%
\alpha}\right\vert ^{2}-\frac{1}{4E_{F}}\left[ \bar{\chi}\sqrt{\bar{\chi}%
^{2}-\Delta ^{2}}\right.  \notag \\
&&\left. -\Delta ^{2}\ln \left( -\frac{\bar{\chi}+\sqrt{\bar{\chi}%
^{2}-\Delta ^{2}}}{\mu -\sqrt{\mu ^{2}+\Delta ^{2}}}\right) \right.  \notag
\\
&&\left. +\mu \left( 2\bar{\chi}-\sqrt{\mu ^{2}+\Delta ^{2}}\right) \right] .
\label{SGE4}
\end{eqnarray}%
From Eqs.~(\ref{GE})-(\ref{PDE}) and (\ref{MPE}), we obtain%
\begin{widetext}
\begin{eqnarray}
\Delta \ln \left( \frac{\sqrt{\Delta ^{2}+\mu ^{2}}-\mu }{E_{b}}\right)
+\Delta \ln \left[ \frac{\left( \bar{\chi}+\sqrt{\bar{\chi}^{2}-\Delta ^{2}}%
\right) \left( \sqrt{\Delta ^{2}+\mu ^{2}}+\mu \right) }{\Delta ^{2}}\right]
&=&0,  \label{D11} \\
\mu +\bar{\chi} &=&2E_{F},  \label{D22} \\
\left[ -\frac{\omega ^{2}\eta ^{2}}{E_{F}\left( \omega ^{2}+\kappa
^{2}\right) }\frac{\mu +\sqrt{\bar{\chi}^{2}-\Delta ^{2}}}{\bar{\chi}}%
+\omega \right] \left\vert \bar{\alpha}\right\vert &=&0,  \label{D33} \\
\frac{\left( \sqrt{\bar{\chi}^{2}-\Delta ^{2}}+\mu \right) }{\bar{\chi}} &=&%
\frac{2E_{F}\bar{m}}{\omega _{0}}.  \label{D44}
\end{eqnarray}%
\end{widetext}By further solving Eqs.~(\ref{D11})-(\ref{D44}), we obtain%
\begin{equation}
\Delta =0\text{, \ \ }\mu =2E_{F}-\omega _{0}\text{, \ \ }\left\vert \bar{%
\alpha}\right\vert =0\text{, \ \ }\bar{m}=1,  \label{S1S}
\end{equation}%
or%
\begin{equation}
\left\{
\begin{array}{l}
\Delta =0, \\
\mu =2\left( E_{F}-\frac{\omega \eta ^{2}}{\omega ^{2}+\kappa ^{2}}\right) ,
\\
\left\vert \bar{\alpha}\right\vert =\sqrt{\frac{\eta ^{2}}{\omega
^{2}+\kappa ^{2}}-\frac{\omega _{0}^{2}\left( \omega ^{2}+\kappa ^{2}\right)
}{4\omega ^{2}\eta ^{2}}}, \\
\bar{m}=\frac{\omega _{0}\left( \omega ^{2}+\kappa ^{2}\right) }{2\omega
\eta ^{2}},%
\end{array}%
\right.  \label{S2S}
\end{equation}%
or%
\begin{equation}
\left\{
\begin{array}{l}
\Delta =\sqrt{E_{b}\left( 2\omega _{0}-E_{b}\right) }, \\
\mu =2E_{F}-\omega _{0}, \\
\left\vert \bar{\alpha}\right\vert =0, \\
\bar{m}=\frac{E_{b}-2\omega _{0}+2E_{F}}{2E_{F}},%
\end{array}%
\right.  \label{S3S}
\end{equation}%
or%
\begin{equation}
\left\{
\begin{array}{l}
\Delta =\sqrt{E_{b}\left[ \frac{2\omega \eta ^{2}\left( 2E_{F}+E_{b}\right)
}{2\omega \eta ^{2}+\left( \omega ^{2}+\kappa ^{2}\right) E_{F}}-E_{b}\right]
}, \\
\mu =2E_{F}-\frac{\omega \eta ^{2}\left( 2E_{F}+E_{b}\right) }{2\omega \eta
^{2}+\left( \omega ^{2}+\kappa ^{2}\right) E_{F}}, \\
\left\vert \bar{\alpha}\right\vert =\frac{\sqrt{\left( \omega ^{2}+\kappa
^{2}\right) }}{2}\sqrt{\left[ \frac{\eta \left( 2E_{F}+E_{b}\right) }{%
2\omega \eta ^{2}+\left( \omega ^{2}+\kappa ^{2}\right) E_{F}}\right] ^{2}-%
\frac{\omega _{0}^{2}}{\omega ^{2}\eta ^{2}}}, \\
\bar{m}=\frac{\omega _{0}\left( \omega ^{2}+\kappa ^{2}\right) }{2\omega
\eta ^{2}}.%
\end{array}%
\right.  \label{S4S}
\end{equation}

Since here the system has two dependent order parameters $\Delta $ and $%
\left\vert \bar{\alpha}\right\vert $, the ground-state stability should be
determined by a $2\times 2$ Hessian matrix \cite{SB81}, which is defined as%
\begin{equation}
M=\left[
\begin{array}{cc}
\frac{\partial ^{2}\bar{E}_{\text{G}}}{\partial \Delta ^{2}} & \frac{%
\partial ^{2}\bar{E}_{\text{G}}}{\partial \Delta \partial \left\vert \bar{%
\alpha}\right\vert } \\
\frac{\partial ^{2}\bar{E}_{\text{G}}}{\partial \left\vert \bar{\alpha}%
\right\vert \partial \Delta } & \frac{\partial ^{2}\bar{E}_{\text{G}}}{%
\partial \left( \left\vert \bar{\alpha}\right\vert \right) ^{2}}%
\end{array}%
\right] .  \label{HM}
\end{equation}%
If $M$ is positive definite (i.e., two eigenvalues of $M$ are positive), $%
\bar{E}_{\text{G}}$ has local minima and the system is located at the stable
phase. If $M$ is indefinite (i.e., one eigenvalues is positive, while the
other is negative), $\bar{E}_{\text{G}}$ has saddle points and the system is
dynamically unstable. If $M$ is negative definite (i.e., two eigenvalues of $%
M$ are negative), $\bar{E}_{\text{G}}$ has a local maximum and the system is
extremely unstable.

In terms of the stability condition given by the Hessian matrix (\ref{HM}),
the ground states corresponding to the solutions (\ref{S3S}) or (\ref{S4S})
are unstable, whereas for the solutions (\ref{S1S}) or (\ref{S2S}) they
become stable. Since $\Delta \equiv 0$ in both Eqs.~(\ref{S1S}) and (\ref%
{S2S}), we can use the similar discussions in the subsection A of Sec.~\ref%
{Phase I}. For example, using the stable condition governed by $\partial ^{2}%
\bar{E}_{\text{G}}/\partial \left( \left\vert \bar{\alpha}\right\vert
\right) ^{2}>0$, we obtain the superradiant critical point $\eta _{c}^{(1)}$%
, which separates the solutions (\ref{S1S}) and (\ref{S2S}). In addition,
the restrictive conditions in Eqs.~(\ref{TH3}) and (\ref{TH3A}) lead to
another critical point $\eta _{c}^{(2)}$. Comparing $\eta _{c}^{(1)}$ with $%
\eta _{c}^{(2)}$, we find that when $\omega _{0}\geqslant E_{F}$, i.e., $%
\eta _{c}^{(1)}>\eta _{c}^{(2)}$, $\bar{\mu}_{-}<0$\ and $\bar{\chi}%
^{2}-\Delta ^{2}\geqslant 0$, and thus, $\Delta $, $\mu $, $\left\vert \bar{%
\alpha}\right\vert $, and $\bar{m}$ are governed by Eq.~(\ref{S1S}) for $%
0<\eta <\eta _{c}^{(1)}$, and for $\eta >\eta _{c}^{(1)}$, they are governed
by Eq.~(\ref{S2S}). When $\omega _{0}<E_{F}$, i.e., $\eta _{c}^{(1)}<\eta
_{c}^{(2)}$, $\bar{\mu}_{-}\geqslant 0$\ and $\bar{\chi}^{2}-\Delta
^{2}\geqslant 0$,\ and thus, for $0<\eta <\eta _{c}^{(2)}$, the scaled
ground-state energy changes and we will discuss the relevant results in the
subsection D of this section. However, for $\eta >\eta _{c}^{(2)}$, $\bar{\mu%
}_{-}<0$\ and $\bar{\chi}^{2}-\Delta ^{2}\geqslant 0$, and thus,$\ \Delta $,
$\mu $, $\left\vert \bar{\alpha}\right\vert $, and $\bar{m}$\ are still
governed by Eq.~(\ref{S2S}).
\begin{figure}[t]
\centering
\includegraphics[width=8cm]{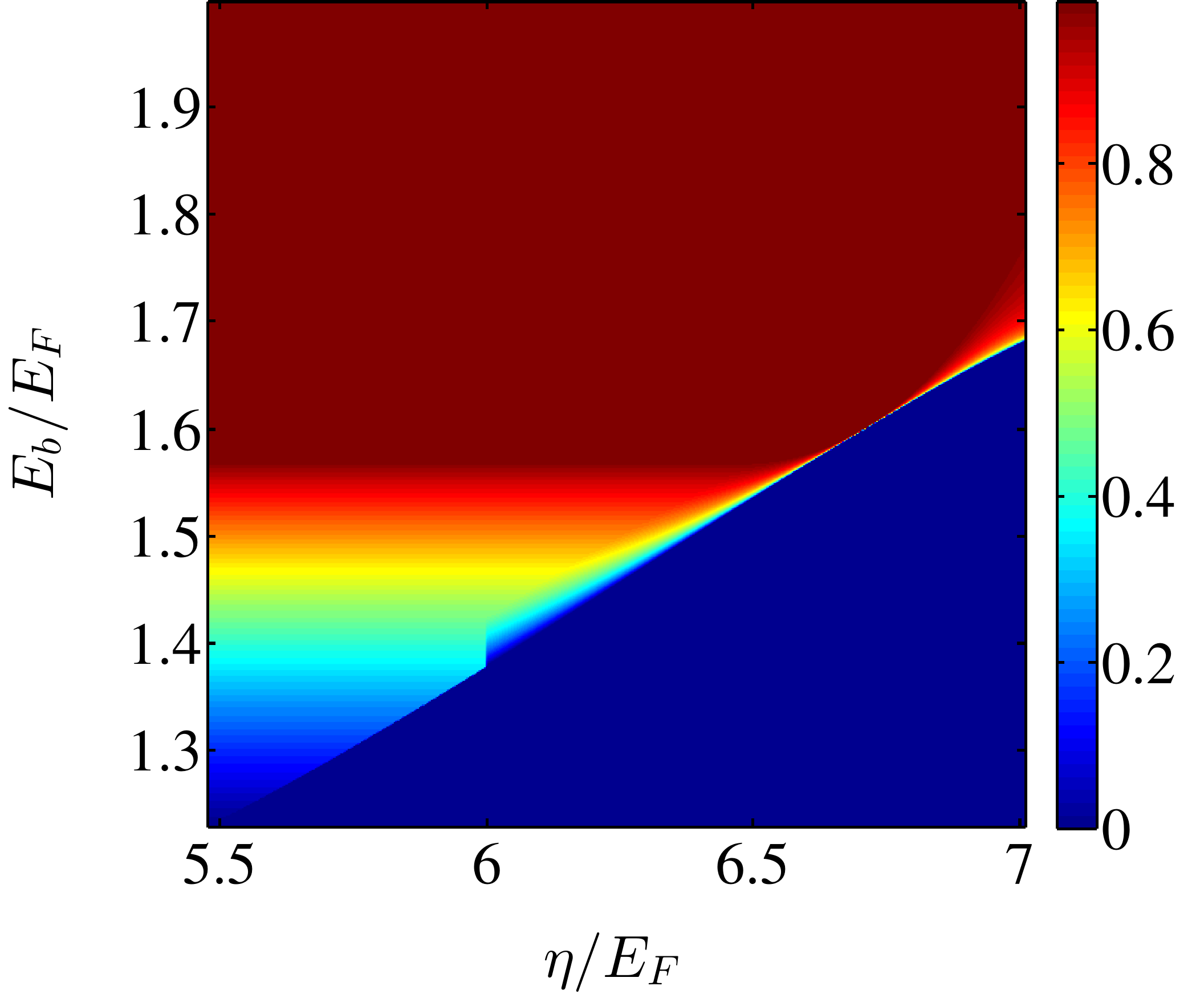}
\caption{The fraction of the SF phase, $x_{2}$, as a function of the
two-body binding energy $E_{b}/E_{F}$ and the effective fermion-photon
coupling strength $\protect\eta /E_{F}$, when the effective resonant
frequency is chosen as $\protect\omega _{0}=1.2E_{F}$. The atom-number
dependent cavity frequency $\protect\omega $ and the cavity decay rate $%
\protect\kappa $ are the same as those in Fig.~\protect\ref{figMP}.}
\label{figfx}
\end{figure}

\subsection{$\bar{\protect\mu}_{-}<0$ and $\bar{\protect\chi}^{2}-\Delta
^{2}<0$}

When%
\begin{equation}
\bar{\mu}_{-}=\mu -\sqrt{\bar{\chi}^{2}-\Delta ^{2}}<0  \label{T3A}
\end{equation}%
and
\begin{equation}
\bar{\chi}^{2}-\Delta ^{2}<0,  \label{TW3}
\end{equation}%
$\Theta \left( \bar{\mu}_{-}\right) =0$ and $\Theta \left( \bar{\chi}%
^{2}-\Delta ^{2}\right) =0$. Thus, the scaled ground-state energy in Eq.~(%
\ref{SGEE4}) becomes
\begin{equation}
\bar{E}_{\text{G}}=\bar{E}_{\text{G}}^{\text{SF}}+\omega \left\vert \bar{%
\alpha}\right\vert ^{2}.  \label{SGEF}
\end{equation}%
From Eqs.~(\ref{GE})-(\ref{PDE}) and (\ref{MPE}), we obtain%
\begin{eqnarray}
\Delta \ln \left( \frac{\sqrt{\Delta ^{2}+\mu ^{2}}-\mu }{E_{b}}\right) &=&0,
\label{D1} \\
\sqrt{\Delta ^{2}+\mu ^{2}}+\mu &=&2E_{F},  \label{D2} \\
2\omega \left\vert \bar{\alpha}\right\vert &=&0,  \label{D3} \\
\bar{m} &=&0.  \label{D4}
\end{eqnarray}%
By further solving Eqs.~(\ref{D1})-(\ref{D4}), we obtain%
\begin{equation}
\Delta =0\text{, \ \ \ \ }\mu =E_{F}\text{, \ \ \ }\left\vert \bar{\alpha}%
\right\vert =0\text{, \ \ \ \ }\bar{m}=0,  \label{SD1}
\end{equation}%
or%
\begin{equation}
\Delta =\sqrt{2E_{F}E_{b}}\text{, \ }\mu =E_{F}-\frac{E_{b}}{2}\text{, }%
\left\vert \bar{\alpha}\right\vert =0\text{, \ }\bar{m}=0.  \label{SD2}
\end{equation}%
Since $\left\vert \bar{\alpha}\right\vert \equiv 0$ in Eqs.~(\ref{SD1}) and (%
\ref{SD2}), we should introduce the stable condition governed by $\partial
^{2}\bar{E}_{\text{G}}/\partial \Delta ^{2}>0$ to find the stable ground
state. According to this stable condition and the restrictive condition in
Eqs.~(\ref{T3A}) and (\ref{TW3}), we find that the ground state, with the
solution (\ref{SD2}), is stable for all $E_{b}$ and $\eta $.

\subsection{The stable ground states for $\protect\omega _{0}\geqslant E_{F}$%
}

In terms of the above discussions in the subsections A and B of this
section, we can obtain the stable ground-state properties for $\omega
_{0}\geqslant E_{F}$. In this case, there exist two kinds of competition
governed by the solutions (\ref{S1S}), (\ref{S2S}), and (\ref{SD2}). When $%
0<\eta <\eta _{c}^{\left( 1\right) }$, the solutions (\ref{S1S}) and (\ref%
{SD2}) dominates, whereas when $\eta >\eta _{c}^{\left( 1\right) }$, the
solutions (\ref{S2S}) and (\ref{SD2}) dominates. These solutions show two
typical properties of the scaled ground-state energy. The first is that the
scaled ground-state energy has a global minimum, i.e., the system is located
at the N-I, SF, or SR phases. The other is that the scaled ground-state
energy has two degenerate minima, which implies that two of these phases can
coexist. Thus, for $\omega _{0}\geqslant E_{F}$, the results for the stable
ground state are summarized as the following two situations: $0<\eta <\eta
_{c}^{\left( 1\right) }$\ and $\eta >\eta _{c}^{\left( 1\right) }$.

\subsubsection{$0<\protect\eta <\protect\eta _{c}^{\left( 1\right) }$}

When $0<\eta <\eta _{c}^{\left( 1\right) }$, it can be seen from Eqs.~(\ref%
{S1S}), (\ref{S2S}), and (\ref{SD2}) that the weak fermion-photon
interaction has no effect on the systematic properties. In this case, only
the N-I and SF phases can be found. More interestingly, when varying $E_{b}$%
, the ground-state energies for these two phases are equal, i.e., these two
phases coexist and the corresponding phase is called the N-I-SF mixed phase.

From the phase equilibrium condition \cite{DES07,LH08}, we find that for $%
E_{b}\!<\!2\left[ \omega _{0}-\left( \sqrt{2}-1\right) E_{F}\right] $, $%
\Delta =0$, and $\bar{E}_{\text{G}}$, $\mu $, and $\bar{m}$ are governed by
Eqs.~(\ref{GN1}) and (\ref{NIG}). This implies that the system is located at
the N-I phase. For $2\left[ \omega _{0}-\left( \sqrt{2}-1\right) E_{F}\right]
<E_{b}<2\!\left[ \omega _{0}-\left( 2-\sqrt{2}\right) E_{F}\right] $, we
find $\bar{E}_{\text{G}}\left( \mu \text{, }\Delta \neq 0\right) =\bar{E}_{%
\text{G}}\left( \mu \text{, }\Delta =0\right) $, which implies that the
system is located at the N-I-SF mixed phase. In order to fully describe the
fundamental properties of this mixed phase, we should introduce the
fractions of the N-I and SF phases, $x_{1}$ and $1-x_{1}$. Moreover, we
further obtain
\begin{eqnarray}
\bar{E}_{\text{G}} &=&\mu -\frac{x_{1}}{4E_{F}}\left( \frac{2\omega
_{0}-E_{b}}{2-\sqrt{2}}\right) ^{2}  \notag \\
&&-\frac{1-x_{1}}{4E_{F}}\left( \frac{2\omega _{0}-E_{b}}{2-\sqrt{2}}\right)
^{2},  \label{GCPP} \\
\Delta  &=&\sqrt{\frac{2\omega _{0}E_{b}-E_{b}^{2}}{\sqrt{2}-1}},
\label{CPTT} \\
\mu  &=&\frac{\sqrt{2}\omega _{0}-E_{b}}{2-\sqrt{2}},  \label{CPMM} \\
\bar{m} &=&1-x_{1},  \label{CPUU}
\end{eqnarray}%
where $x_{1}=2\sqrt{2}E_{F}/\left( 2\omega _{0}-E_{b}\right) -\sqrt{2}-1$.
The detailed derivation of the above results is given by the Appendix~\ref%
{AppAII}. For $E_{b}>2\!\left[ \omega _{0}-\left( 2-\sqrt{2}\right) E_{F}%
\right] $, we find
\begin{equation}
\bar{E}_{\text{G}}=-\frac{E_{F}}{2}\text{, \ }\Delta =\sqrt{2E_{F}E_{b}}%
\text{, }\mu =E_{F}-\frac{E_{b}}{2}\text{, }\bar{m}=0,  \label{GCPR}
\end{equation}%
which indicates that the system is located at the SF phase. The analytical
results in Eqs.~(\ref{GN1}), (\ref{NIG}), and (\ref{GCPP})-(\ref{GCPR}) are
the same as those in Refs.~\cite{LH08,DE15}, as expected. They show that
when increasing $E_{b}$, two first-order phase transitions from the N-I
phase to the N-I-SF mixed phase or from the N-I-SF mixed phase to the SF
phase emerge \cite{DES06,HC12,LH08,DE15,KBG06,PFB03,GBP06,MWZ06}. Moreover,
the ratio of the scaled polarization to the dimensionless mean-field gap in
the N-II-SF mixed phase, $\bar{m}/\left( \Delta /E_{F}\right) $, is
decreased.

\begin{figure}[t]
\centering
\includegraphics[width=7cm]{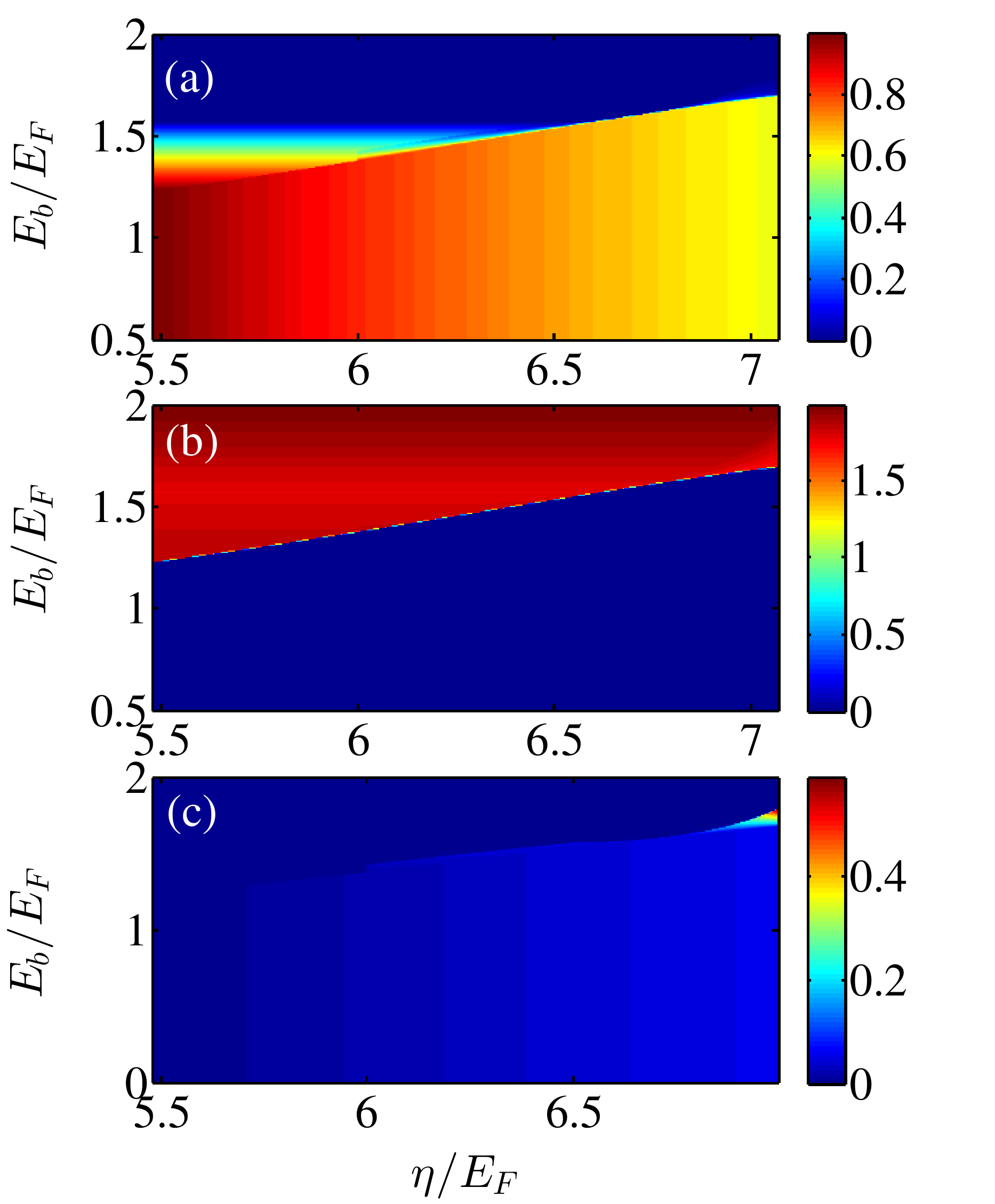}
\caption{(a) The scaled polarization $\bar{m}$, (b) the mean-field gap $%
\Delta /E_{F}$, and (c) the scaled mean-photon number $\left\vert \bar{%
\protect\alpha}\right\vert ^{2}$ as functions of the two-body binding energy
$E_{b}/E_{F}$ and the effective fermion-photon coupling strength $\protect%
\eta /E_{F}$, when the effective resonant frequency is chosen as $\protect%
\omega _{0}=1.2E_{F}$. The atom-number dependent cavity frequency $\protect%
\omega $ and the cavity decay rate $\protect\kappa $ are the same as those
in Fig.~\protect\ref{figMP}.}
\label{ODSSB}
\end{figure}

\subsubsection{$\protect\eta >\protect\eta _{c}^{\left( 1\right) }$}

When $\eta >\eta _{c}^{\left( 1\right) }$, it can be seen from Eq.~(\ref{S2S}%
) that a non-zero $\left\vert \bar{\alpha}\right\vert $ emerges, which means
that the fermion-photon interaction has a significant effect on the
systematic properties. In this case, only the SF and SR phases can be found.
More interestingly, when varying $E_{b}$ and $\eta $, the ground-state
energies for these two phases are equal, i.e., these two phases coexist and
the corresponding phase is called the \textit{SF-SR mixed phase}.

\begin{figure}[t]
\centering
\includegraphics[width=8.5cm]{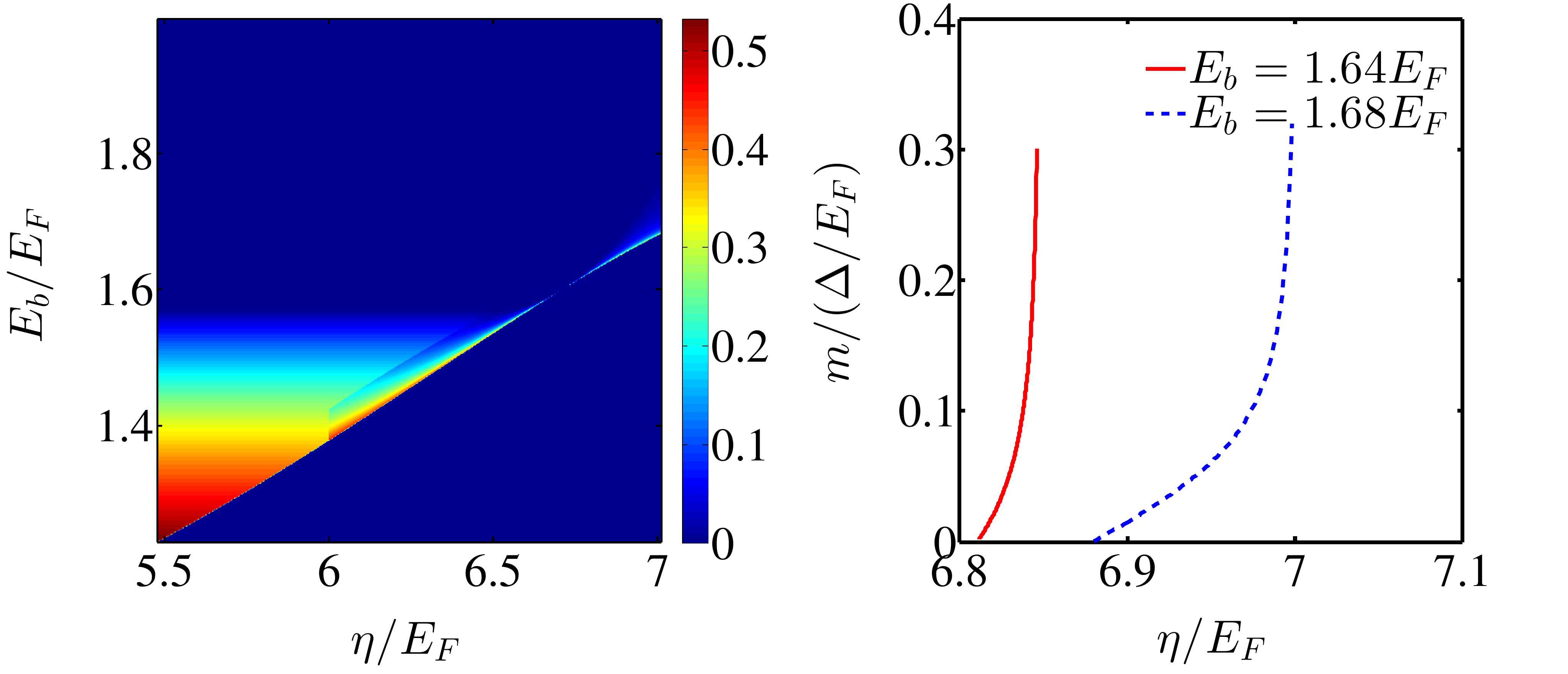}
\caption{(a) The ratio of the scaled polarization to the dimensionless
mean-field gap in the N-II-SF mixed state, $\bar{m}/\left( \Delta
/E_{F}\right) $, as a function of the two-body binding energy $E_{b}/E_{F}$
and the effective fermion-photon coupling strength $\protect\eta /E_{F}$,
when the effective resonant frequency is chosen as $\protect\omega %
_{0}=1.2E_{F}$. The atom-number dependent cavity frequency $\protect\omega $
and the cavity decay rate $\protect\kappa $ are the same as those in Fig.~%
\protect\ref{figMP}. (b) For a fixed $E_{b}=E_{F}$ (the red-solid line) and $%
E_{b}=1.1E_{F}$ (the blue-dashed line), $\bar{m}/\left( \Delta /E_{F}\right)
$ varies as a function of $\protect\eta /E_{F}$.}
\label{figrps}
\end{figure}

From the phase equilibrium condition \cite{DES07,LH08}, we find four stable
regions as follows.

(i) When $E_{b}<E_{b}^{\left( 1\right) }$, $\Delta =0$, and $\bar{E}_{\text{G%
}}$, $\mu $, $\left\vert \bar{\alpha}\right\vert $, and $\bar{m}$ are
governed by Eqs.~(\ref{LAG})-(\ref{LAm}). These mean that the system is
located at the SR phase.

(ii) When $E_{b}^{\left( 1\right) }\Theta \left( \eta _{c}^{\left( 3\right)
}-\eta \right) \!<\!E_{b}\!<\!2\left[ \omega _{0}-\left( 2-\sqrt{2}\right)
E_{F}\right] \Theta \left( \eta _{c}^{\left( 3\right) }-\eta \right) $ or $%
E_{b}^{\left( 2\right) }\Theta \left( \eta -\eta _{c}^{\left( 3\right)
}\right) \!<\!E_{b}\!<\!2\left[ \omega _{0}-\left( 2-\sqrt{2}\right) E_{F}%
\right] \Theta \left( \eta -\eta _{c}^{\left( 3\right) }\right) $, with $%
\eta _{c}^{\left( 3\right) }=\omega _{0}\sqrt{\left( \omega ^{2}+\kappa
^{2}\right) /2\omega }$, we find $\bar{E}_{\text{G}}\left( \mu \text{, }%
\Delta \neq 0\text{, }\left\vert \bar{\alpha}\right\vert =0\right) =\bar{E}_{%
\text{G}}\left( \mu \text{, }\Delta =0\text{, }\left\vert \bar{\alpha}%
\right\vert =0\right) $, which means that the N-I and SF phases coexist and
the corresponding phase is called the N-I-SF mixed phase. We further obtain $%
\bar{E}_{\text{G}}$, $\Delta $, $\mu $, and $\bar{m}$, which are governed by
Eqs.~(\ref{GCPP})-(\ref{CPUU}).

(iii) When $E_{b}^{\left( 1\right) }\!<\!E_{b}\!<\!E_{b}^{\left( 2\right)
}\Theta \left( \eta -\eta _{c}^{\left( 3\right) }\right) $, we find $\bar{E}%
_{\text{G}}\left( \mu \text{, }\Delta \neq 0\text{, }\left\vert \bar{\alpha}%
\right\vert =0\right) =\bar{E}_{\text{G}}\left( \mu \text{, }\Delta =0\text{%
, }\left\vert \bar{\alpha}\right\vert \neq 0\right) $, which implies that
the system is located at the SF-SR mixed phase. In order to fully describe
the fundamental properties of this mixed phase, we should introduce the
fractions of the SF and SR phases, $x_{2}$ and $1-x_{2}$. Moreover, we
further obtain
\begin{eqnarray}
\bar{E}_{\text{G}} &=&\mu -x_{2}\frac{1}{2E_{F}}\left( \mu +\frac{E_{b}}{2}%
\right) ^{2}  \notag \\
&&-\left( 1-x_{2}\right) \left[ \frac{\mu ^{2}}{4E_{F}A}+\frac{\omega
_{0}^{2}\left( \omega ^{2}+\kappa ^{2}\right) }{4\omega \eta ^{2}}\right] ,
\label{GCP2} \\
\Delta &=&\sqrt{E_{b}\left( E_{b}+2\mu \right) },  \label{CPF} \\
\mu &=&\frac{-E_{b}A\pm \sqrt{A^{2}E_{b}^{2}-2A\left( 2A-1\right) B}}{2A-1},
\label{VPC} \\
\left\vert \bar{\alpha}\right\vert &=&\frac{1}{2}\sqrt{\frac{\mu ^{2}\eta
^{2}}{\left( \omega ^{2}+\kappa ^{2}\right) E_{F}^{2}A^{2}}-\frac{\omega
_{0}^{2}\left( \omega ^{2}+\kappa ^{2}\right) }{\omega ^{2}\eta ^{2}}},
\label{CPA} \\
\bar{m} &=&\left( 1-x_{2}\right) \frac{\omega _{0}\left( \omega ^{2}+\kappa
^{2}\right) }{2\omega \eta ^{2}},  \label{CPV}
\end{eqnarray}%
where $A=1-\omega \eta ^{2}/\left[ E_{F}\left( \omega ^{2}+\kappa
^{2}\right) \right] $, $B=E_{b}^{2}/4-E_{F}\omega _{0}^{2}\left( \omega
^{2}+\kappa ^{2}\right) /\left( 2\omega \eta ^{2}\right) $, and $%
x_{2}=\left( 2E_{F}A-\mu \right) /\left( 2\mu A-\mu +E_{b}A\right) $. The
detailed derivation of the above results is displayed in the Appendix~\ref%
{AppAIII}. In principle, $E_{b}^{\left( 1\right) }$ and $E_{b}^{\left(
2\right) }$ can be obtained analytically. However, their expressions are so
complicated that here we donot list them.

The analytical results in Eqs.~(\ref{CPF})-(\ref{CPV}) show that the
predicted SF-SR mixed phase has the following typical properties:

$\bullet $ When $x_{2}=0$, the system is located at the SR phase with $\bar{m%
}=$ $\omega _{0}\left( \omega ^{2}+\kappa ^{2}\right) /\left( 2\omega \eta
^{2}\right) $, whereas when $x_{2}=1$, the system enters into the SF phase
with $\bar{m}=0$. The above explicit expressions show that the nonzero $\bar{%
m}$ in the SF-SR\ mixed phase\ is only caused by the macroscopic collective
excitation of both the fermions and photons, which is different from that of
the N-I-SF phase.

$\bullet $ For a relative small $E_{b}$ or larger $\eta $, $x_{2}\rightarrow
0$, as shown in Fig.~\ref{figfx}. This means that the systematic properties
are mainly governed by the SR properties. Whenever increasing $\eta $ or $%
E_{b}$, $\bar{m}$ is decreased, and $\Delta $ and $\left\vert \bar{\alpha}%
\right\vert $ are increased, as shown in Figs.~\ref{ODSSB}(a)-\ref{ODSSB}(c).

$\bullet $ For a relative small $\eta $ or larger $E_{b}$, $x_{2}\rightarrow
1$, as also shown in Fig.~\ref{figfx}. This means that the systematic
properties are mainly governed by the SF properties. In this case, $\bar{m}$
approaches zero, as also shown in Fig.~\ref{ODSSB}(a), and $\Delta $ and $%
\left\vert \bar{\alpha}\right\vert $ almost reach their maximum values, as
also shown in Figs.~\ref{ODSSB}(b) and \ref{ODSSB}(c).

\begin{figure}[t]
\centering
\includegraphics[width=7cm]{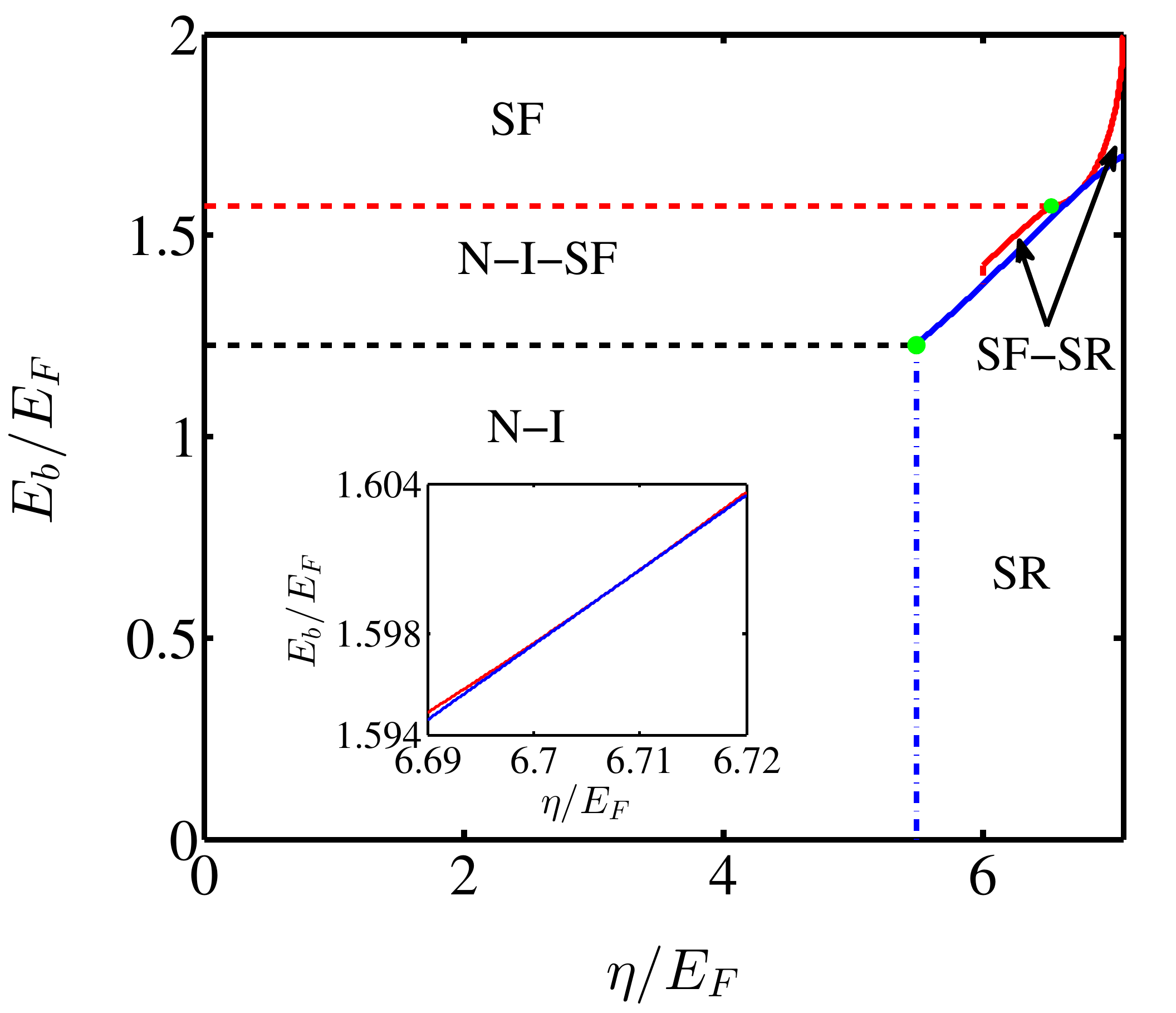}
\caption{Phase diagram as a function of the two-body binding energy $%
E_{b}/E_{F}$ and the effective atom-photon coupling strength $\protect\eta %
/E_{F}$, when the effective resonant frequency is chosen as $\protect\omega %
_{0}=1.2E_{F}$. The atom-number dependent cavity frequency $\protect\omega $
and the cavity decay rate $\protect\kappa $ are also the same as in Fig.~%
\protect\ref{figMP}. Inset: the region of the phase boundaries approaching
each other.}
\label{wphase12}
\end{figure}

$\bullet $ For the intermediate $\eta $ and $E_{b}$, both $x_{2}$ and $%
1-x_{2}$ are the finite values ranging from $0$ to $1$, as also shown in
Fig.~\ref{figfx}. These mean that the SF and SR properties have a strong
competition. When increasing $E_{b}$, $\bar{m}$ is decreased, and thus, both
$\Delta $ and $\left\vert \bar{\alpha}\right\vert $ are increased, as also
shown in Figs.~\ref{ODSSB}(a)-\ref{ODSSB}(c). However, when increasing $\eta
$, $\bar{m}$ is increased, due to the rapid increasing of $1-x_{2}$, i.e.,
the fraction of the SR phase, and thus, both $\Delta $ and $\left\vert \bar{%
\alpha}\right\vert $ are decreased, as also shown in Figs.~\ref{ODSSB}(a)-%
\ref{ODSSB}(c). This is quite different from that in the SR phase, in which
when increasing $\eta $, $\bar{m}$ is decreased [see Fig.~\ref{figMP}(d)].
In order to see clearly the evolution of $\bar{m}$ and $\Delta $, we plot $%
\bar{m}/\left( \Delta /E_{F}\right) $ as a function of $\eta $ and $E_{b}$
in Fig.~\ref{figrps}. For a fixed $E_{b}$, when increasing $\eta $, $\bar{m}%
/\left( \Delta /E_{F}\right) $ is increased, as shown in Figs.~\ref{figrps}%
(a) and \ref{figrps}(b). Based on this conclusion, we expect that in real
experiments we can tune $\eta $ and $E_{b}$ to find a relative large regime
that the magnetic and SF properties coexist. This is also different from the
situation in the N-II-SF mixed phase, in which when increasing $E_{b}$ the
coexisted regime becomes smaller and smaller. In addition, $\bar{m}%
/\left\vert \bar{\alpha}\right\vert $ has a similar behavior, and thus, is
not addressed here.

(iv) When $E_{b}\!>\!2\!\left[ \omega _{0}-\left( 2-\sqrt{2}\right) E_{F}%
\right] \Theta \left( \eta _{c}^{\left( 3\right) }-\eta \right) $ or $%
E_{b}>E_{b}^{\left( 2\right) }\Theta \left( \eta -\eta _{c}^{\left( 3\right)
}\right) $, $\left\vert \bar{\alpha}\right\vert =0$, and $\bar{E}_{\text{G}}$%
, $\Delta $, $\mu $, and $\bar{m}$\ are governed by Eq.~(\ref{GCPR}). These
mean that the system is located at the SF phase.

\subsubsection{Phase diagram}

In Fig.~\ref{wphase12}, we plot the whole phase diagram, including the N-I
phase, the N-I-SR mixed phase, the SF-SR mixed phase, the SF phase, and the
SR phase, for $0<\eta <\eta _{c}^{\left( 1\right) }$ and $\eta >\eta
_{c}^{\left( 1\right) }$. The phase transitions, from the N-I-SF mixed phase
to the N-I phase or the SF phase or the SR phase or the SF-SR mixed phase
and from the SF-SR mixed phase to the SF phase or the SP phase, are of the
first order, due to the existence of the N-I-SF and SF-SR mixed phases.
However, the phase transition from the N-I phase to the SR phase is of the
second order. In addition, this phase diagram has a tricritical point (the
green dot), at which the phase transition changes from the first order to
the second order.

\subsection{$\bar{\protect\mu}_{-}\geqslant 0$ and $\bar{\protect\chi}%
^{2}-\Delta ^{2}\geqslant 0$}

When%
\begin{equation}
\bar{\mu}_{-}=\mu -\sqrt{\bar{\chi}^{2}-\Delta ^{2}}\geqslant 0  \label{TS}
\end{equation}%
and
\begin{equation}
\bar{\chi}^{2}-\Delta ^{2}\geqslant 0,  \label{TS1}
\end{equation}%
$\Theta \left( \bar{\mu}_{-}\right) =1$ and $\Theta \left( \bar{\chi}%
^{2}-\Delta ^{2}\right) =1$. Thus, the scaled ground-state energy in Eq.~(%
\ref{SGEE4}) becomes
\begin{eqnarray}
\bar{E}_{\text{G}} &=&\bar{E}_{\text{G}}^{\text{SF}}+\omega \left\vert \bar{%
\alpha}\right\vert ^{2}-\frac{1}{4E_{F}}\left[ 2\bar{\chi}\sqrt{\bar{\chi}%
^{2}-\Delta ^{2}}\right.  \notag \\
&&\left. -\Delta ^{2}\ln \left( \frac{\bar{\chi}+\sqrt{\bar{\chi}^{2}-\Delta
^{2}}}{\bar{\chi}-\sqrt{\bar{\chi}^{2}-\Delta ^{2}}}\right) \right] .
\label{SGSE}
\end{eqnarray}%
From Eqs.~(\ref{GE})-(\ref{PDE}) and (\ref{MPE}), we obtain%
\begin{widetext}
\begin{eqnarray}
\Delta \ln \left( \frac{\sqrt{\Delta ^{2}+\mu ^{2}}-\mu }{E_{b}}\right)
+\Delta \ln \left[ \frac{2\bar{\chi}\left( \bar{\chi}+\sqrt{\bar{\chi}%
^{2}-\Delta ^{2}}\right) -\Delta ^{2}}{\Delta ^{2}}\right] &=&0,  \label{G1}
\\
\sqrt{\Delta ^{2}+\mu ^{2}}+\mu &=&2E_{F},  \label{C1} \\
\left[ -\frac{2\omega ^{2}\eta ^{2}}{E_{F}\left( \omega ^{2}+\kappa
^{2}\right) }\frac{\sqrt{\bar{\chi}^{2}-\Delta ^{2}}}{\bar{\chi}}+\omega %
\right] \left\vert \bar{\alpha}\right\vert &=&0,  \label{A1} \\
\frac{2\sqrt{\bar{\chi}^{2}-\Delta ^{2}}}{\bar{\chi}} &=&\frac{2E_{F}\bar{m}%
}{\omega _{0}}.  \label{m1}
\end{eqnarray}%
\end{widetext}By further solving Eqs.~(\ref{G1})-(\ref{m1}), we obtain the
following solutions:
\begin{equation}
\Delta =0\text{, \ }\mu =E_{F}\text{, \ \ }\left\vert \bar{\alpha}%
\right\vert =0\text{, \ }\bar{m}=\frac{\omega _{0}}{E_{F}},  \label{s1s}
\end{equation}%
or
\begin{equation}
\left\{
\begin{array}{l}
\Delta =\sqrt{\sqrt{2E_{F}E_{b}}\left( 2\omega _{0}-\sqrt{2E_{F}E_{b}}%
\right) },\text{ } \\
\mu =E_{F}+\frac{E_{b}}{2}-\frac{\omega _{0}\sqrt{2E_{F}E_{b}}}{2E_{F}},%
\text{ } \\
\left\vert \bar{\alpha}\right\vert =0, \\
\bar{m}=\frac{\sqrt{\omega _{0}^{2}-\sqrt{2E_{F}E_{b}}\left( 2\omega _{0}-%
\sqrt{2E_{F}E_{b}}\right) }}{E_{F}},%
\end{array}%
\right.  \label{s2s}
\end{equation}%
or
\begin{equation}
\left\{
\begin{array}{l}
\Delta =\sqrt{2E_{F}E_{b}}\sqrt{\frac{2\omega \eta ^{2}-E_{F}\left( \omega
^{2}+\kappa ^{2}\right) }{2\omega \eta ^{2}+E_{F}\left( \omega ^{2}+\kappa
^{2}\right) }}, \\
\mu =E_{F}+\frac{E_{b}}{2}-\frac{2\omega \eta ^{2}E_{b}}{2\omega \eta
^{2}+E_{F}\left( \omega ^{2}+\kappa ^{2}\right) }, \\
\left\vert \bar{\alpha}\right\vert =\sqrt{\frac{2\eta ^{2}E_{F}E_{b}\left(
\omega ^{2}+\kappa ^{2}\right) }{\left[ 2\omega \eta ^{2}+E_{F}\left( \omega
^{2}+\kappa ^{2}\right) \right] ^{2}}-\frac{\omega _{0}^{2}\left( \omega
^{2}+\kappa ^{2}\right) }{4\omega ^{2}\eta ^{2}}}, \\
\bar{m}=\frac{\omega _{0}\left( \omega ^{2}+\kappa ^{2}\right) }{2\omega
\eta ^{2}}.%
\end{array}%
\right.  \label{s3s}
\end{equation}

In terms of the stability condition given by the Hessian matrix (\ref{HM}),
the ground states corresponding to the solutions (\ref{s2s}) and (\ref{s3s})
are unstable, whereas for the solution (\ref{s1s}) it becomes stable.\ Since
$\Delta \equiv 0$\ in Eq.~(\ref{s1s}), we can use the similar discussions in
the subsection B of Sec.~\ref{Phase I}. For instance, using the stable
condition governed by $\partial ^{2}\bar{E}_{\text{G}}/\partial \left(
\left\vert \bar{\alpha}\right\vert \right) ^{2}>0$ and the restrictive
conditions in Eqs.~(\ref{TS}) and (\ref{TS1}), we find that when $\omega
_{0}<E_{F}$, $\bar{\mu}_{-}\geqslant 0$ and $\bar{\chi}^{2}-\Delta
^{2}\geqslant 0$, and thus, for $0<\eta <\eta _{c}^{(2)}$, $\Delta $, $\mu $%
, $\left\vert \bar{\alpha}\right\vert $, and $\bar{m}$\ are governed by Eq.~(%
\ref{s1s}), whereas for $\eta >\eta _{c}^{(2)}$, $\bar{\mu}_{-}<0$ and $\bar{%
\chi}^{2}-\Delta ^{2}\geqslant 0$,\ we should combine with the previous
discussions in the subsection A of this section, and thus, $\Delta $, $\mu $%
, $\left\vert \bar{\alpha}\right\vert $, and $\bar{m}$\ are governed by Eq.~(%
\ref{S2S}).

\subsection{$\bar{\protect\mu}_{-}\geqslant 0$ and $\bar{\protect\chi}%
^{2}-\Delta ^{2}<0$}

When%
\begin{equation}
\bar{\mu}_{-}=\mu -\sqrt{\bar{\chi}^{2}-\Delta ^{2}}\geqslant 0  \label{TC}
\end{equation}%
and
\begin{equation}
\bar{\chi}^{2}-\Delta ^{2}<0,  \label{TC1}
\end{equation}%
$\Theta \left( \bar{\mu}_{-}\right) =1$ and $\Theta \left( \bar{\chi}%
^{2}-\Delta ^{2}\right) =0$. The scaled ground-state energy in Eq.~(\ref%
{SGEE4}) becomes
\begin{equation}
\bar{E}_{\text{G}}=\bar{E}_{\text{G}}^{\text{SF}}+\omega \left\vert \bar{%
\alpha}\right\vert ^{2},  \label{SGFF}
\end{equation}%
which is the same as Eq.~(\ref{SGEF}). Thus, the stable ground-state
properties are the same with those in the subsection B of this section.

\subsection{The stable ground states for $\protect\omega _{0}<E_{F}$}

In terms of the above discussions in the subsections A, D, and E of this
section, we can obtain the stable ground-state properties for $\omega
_{0}<E_{F}$. In this case, there also exist two kinds of competition
governed by the solutions (\ref{S2S}), (\ref{SD2}), and (\ref{s1s}). When $%
0<\eta <\eta _{c}^{\left( 2\right) }$, the solutions (\ref{SD2}) and (\ref%
{s1s}) dominates, whereas when $\eta >\eta _{c}^{\left( 2\right) }$, the
solutions (\ref{S2S}) and (\ref{SD2}) dominates. These solutions also show
two typical properties of the scaled ground-state energy. The first is that
the scaled ground-state energy has a global minimum, i.e., the system is
located at the N-II, SF, and SR phases. The other is that the scaled
ground-state energy has two degenerate minima, which implies that two of
these phases can coexist. Thus, for $\omega _{0}<E_{F}$, the results for the
stable ground state are summarized as the following two situations: $0<\eta
<\eta _{c}^{\left( 2\right) }$\ and $\eta >\eta _{c}^{\left( 2\right) }$.

\begin{figure}[!t]
\centering
\includegraphics[width=7cm]{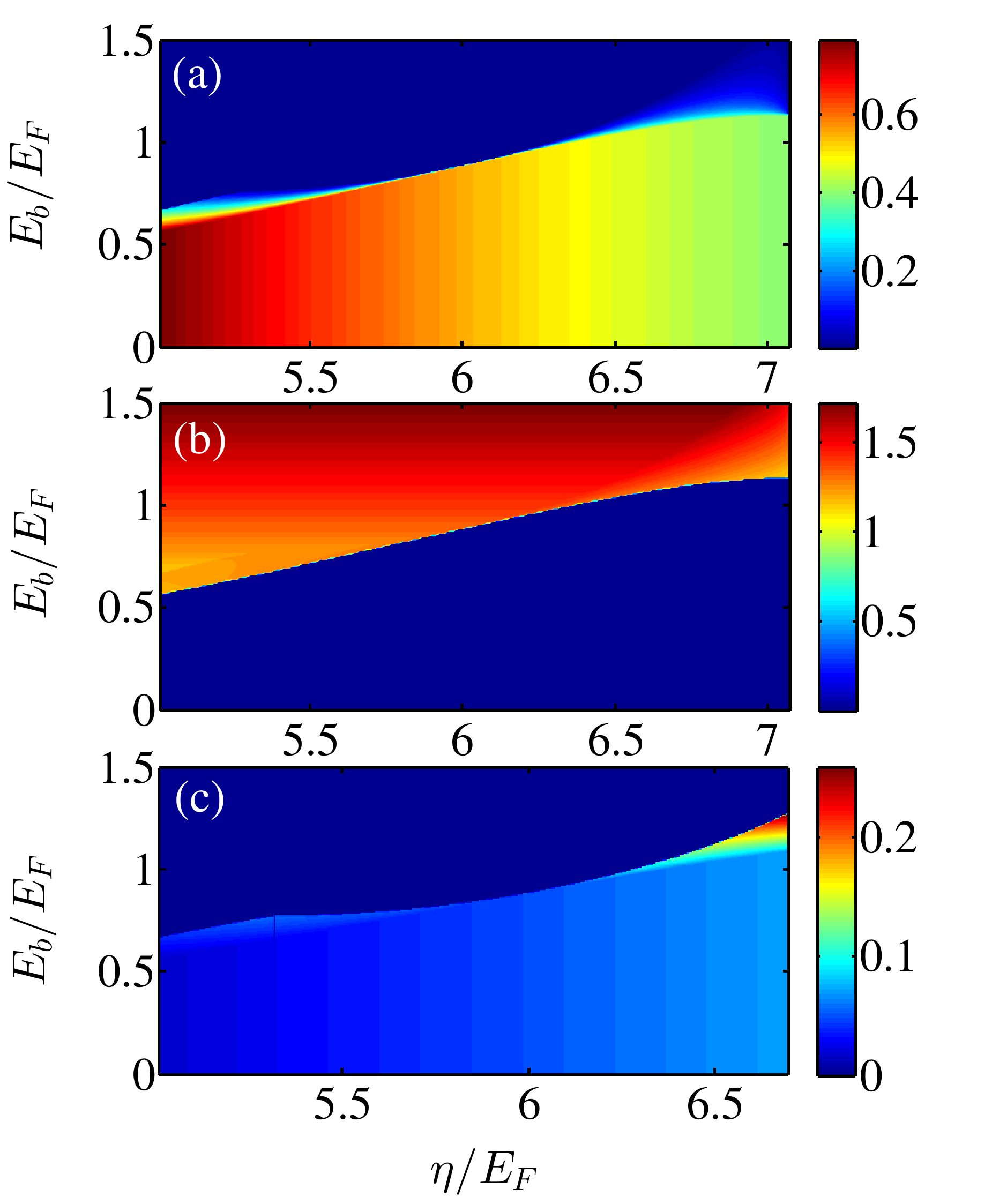}
\caption{(a) The scaled polarization $\bar{m}$, (b) the mean-field gap $%
\Delta /E_{F}$, and (c) the scaled mean-photon number $\left\vert \bar{%
\protect\alpha}\right\vert ^{2}$ as functions of the two-body binding energy
$E_{b}/E_{F}$ and the effective fermion-photon coupling strength $\protect%
\eta /E_{F}$, when the effective resonant frequency is chosen as $\protect%
\omega _{0}=0.8E_{F}$. The atom-number dependent cavity frequency $\protect%
\omega $ and the cavity decay rate $\protect\kappa $ are the same as those
in Fig.~\protect\ref{figMP}.}
\label{ODSS}
\end{figure}

\subsubsection{$0<\protect\eta <\protect\eta _{c}^{\left( 2\right) }$}

When $0<\eta <\eta _{c}^{\left( 2\right) }$, it can be seen from Eqs.~(\ref%
{S2S}), (\ref{SD2}), and (\ref{s1s}) that the weak fermion-photon
interaction has no effect on the systematic properties. In this case, only
the N-II and SF phases can be found. More interestingly, when varying $E_{b}$%
, the ground-state energies for these two phases are equal, i.e., these two
phases coexist and the corresponding phase is called the N-II-SF mixed phase.

From the phase equilibrium condition \cite{DES07,LH08}, we find that for $%
E_{b}<2\left( \sqrt{E_{F}^{2}+\omega _{0}^{2}}-E_{F}\right) $, $\Delta =0$,
and $\bar{E}_{\text{G}}$, $\mu $, and $\bar{m}$ are governed by Eqs.~(\ref%
{GN2}) and (\ref{NTG}). This implies that the system is located at the N-II
phase. For $2\!\left( \!\sqrt{E_{F}^{2}+\omega _{0}^{2}}-E_{F}\right)
\!<\!E_{b}\!<\!2\left( \!E_{F}-\sqrt{E_{F}^{2}-\omega _{0}^{2}}\right)
\Theta \left( \omega _{0}-\omega _{01}\right) $, where $\omega
_{01}=E_{b}\left( 1+\sqrt{2}\right) /2$ is determined by $\mu _{c1}=\omega
_{0}$, we find $\bar{E}_{\text{G}}\left( \mu \text{, }\Delta =0\right) =\bar{%
E}_{\text{G}}\left( \mu \text{, }\Delta \neq 0\right) $. This implies that
the system is located at the N-II-SF mixed phase. In order to fully describe
the fundamental properties of this mixed phase, we should introduce the
fractions of the N-II and SF phases, $x_{3}$ and $1-x_{3}$. Moreover, we
further obtain%
\begin{eqnarray}
\bar{E}_{\text{G}} &=&\mu -\frac{x_{3}}{2E_{F}}\left( \frac{\omega _{0}^{2}}{%
E_{b}}+\frac{E_{b}}{4}\right) ^{2}  \notag \\
&&-\left( 1-x_{3}\right) \frac{\left( 4\omega _{0}^{2}+E_{b}^{2}\right) ^{2}%
}{32E_{F}E_{b}^{2}},  \label{GCP} \\
\Delta &=&\sqrt{\frac{E_{b}^{2}+4\omega _{0}^{2}}{2}},  \label{CPT} \\
\mu &=&\frac{\omega _{0}^{2}}{E_{b}}-\frac{E_{b}}{4},  \label{CPM} \\
\bar{m} &=&\left( 1-x_{3}\right) \frac{\omega _{0}}{E_{F}},  \label{CPU}
\end{eqnarray}%
where $x_{3}=1/2+2E_{F}/E_{b}-2\omega _{0}^{2}/E_{b}^{2}$. The detailed
derivation of the above results is given by the Appendix~\ref{AppAII}. For $%
E_{b}>2\left( E_{F}-\sqrt{E_{F}^{2}-\omega _{0}^{2}}\right) \Theta \left(
\omega _{01}-\omega _{0}\right) $, we find that $\bar{E}_{\text{G}}$, $%
\Delta $, $\mu $, and $\bar{m}$ are governed by Eq.~(\ref{GCPR}), which also
indicates that the system is located at the SF phase. The analytical results
in Eqs.~(\ref{GN2}), (\ref{NTG}), (\ref{GCPR}), and (\ref{GCP})-(\ref{CPU})
are also the same as those in Refs.~\cite{LH08,DE15}, as expected. The basic
properties of the N-II-SF mixed phase are similar to those in the N-I-SF
mixed phase, and thus, are not discussed here.

\subsubsection{$\protect\eta >\protect\eta _{c}^{\left( 2\right) }$}

When $\eta >\eta _{c}^{\left( 2\right) }$, it can be seen from Eq.~(\ref{S2S}%
) that the fermion-photon interaction plays a significant role in the
systematic properties, which are sharply contrast to the case of $0<\eta
<\eta _{c}^{\left( 2\right) }$ and similar to the case of $\eta >\eta
_{c}^{\left( 1\right) }$ in the subsubsection 2 of this section. In terms of
Eqs.~(\ref{S2S}) and (\ref{SD2}), we plot $\Delta $, $\mu $, $\left\vert
\bar{\alpha}\right\vert $, and $\bar{m}$ as functions of $E_{b}$ and $\eta $
in Fig.~\ref{ODSS}, and find three stable regions as follows.

(i) When $E_{b}<E_{b}^{\left( 1\right) }$, $\Delta =0$, and $\bar{E}_{\text{G%
}}$, $\mu $, $\left\vert \bar{\alpha}\right\vert $, and $\bar{m}$ are
governed by Eqs.~(\ref{LAM}) and (\ref{Cs}). These mean that the system is
located at the SR phase.

(ii) When $E_{b}^{\left( 1\right) }\!<\!E_{b}\!<\!E_{b}^{\left( 2\right) }$,
we find $\bar{E}_{\text{G}}\left( \mu \text{, }\Delta \neq 0\text{, }%
\left\vert \bar{\alpha}\right\vert =0\right) =\bar{E}_{\text{G}}\left( \mu
\text{, }\Delta =0\text{, }\left\vert \bar{\alpha}\right\vert \neq 0\right) $%
, which means that the SF and SR phases coexist and the corresponding phase
is called the SF-SR\ mixed phase. We further obtain $\bar{E}_{\text{G}}$, $%
\Delta $, $\mu $, $\left\vert \bar{\alpha}\right\vert $, and $\bar{m}$,
which are governed by Eqs.~(\ref{GCP2})-(\ref{CPV}). The other typical
properties in this SF-SR mixed phase are the same as those in the region of $%
\eta \geqslant \eta _{c}^{\left( 1\right) }$, and thus, are not addressed
here.

\begin{figure}[t]
\centering
\includegraphics[width=7cm]{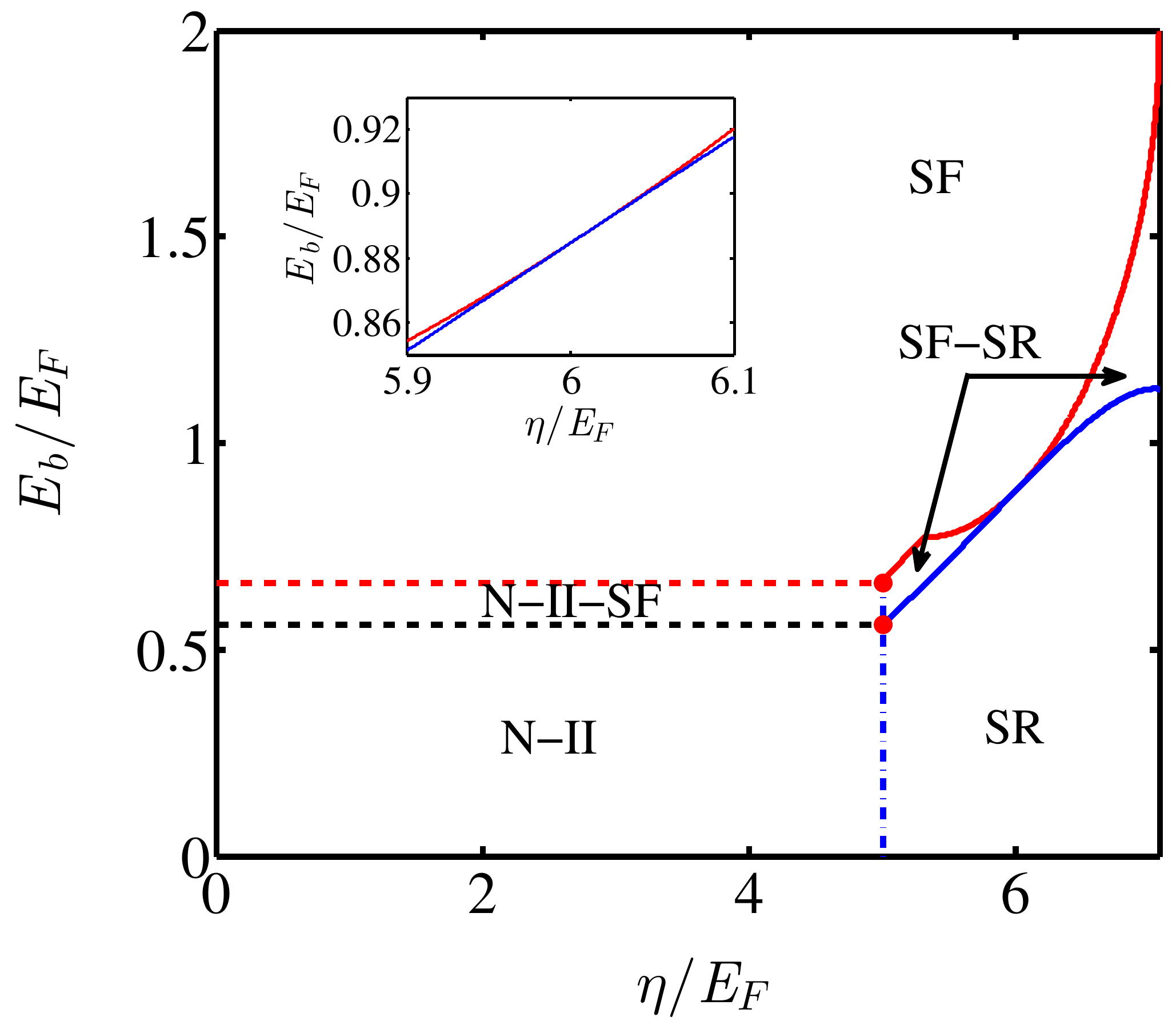}
\caption{Phase diagram as a function of the two-body binding energy $%
E_{b}/E_{F}$ and the effective atom-photon coupling strength $\protect\eta %
/E_{F}$, when the effective resonant frequency is chosen as $\protect\omega %
_{0}=0.8E_{F}$. The atom-number dependent cavity frequency $\protect\omega $
and the cavity decay rate $\protect\kappa $ are also same as those in Fig.~%
\protect\ref{figMP}. Inset: the region of the phase boundaries approaching
each other.}
\label{wphase08}
\end{figure}

(iii) When $E_{b}>E_{b}^{\left( 2\right) }$, $\left\vert \bar{\alpha}%
\right\vert =0$, and $\bar{E}_{\text{G}}$, $\Delta $, $\mu $, and $\bar{m}$
are governed by Eq.~(\ref{GCPR}). These mean that the system is located at
the SF phase.

\subsubsection{Phase diagram}

In Fig.~\ref{wphase08}, we plot the whole phase diagram, including the N-II
phase, the N-II-SR mixed phase, the SF-SR mixed phase, the SF phase, and the
SR phase, for $0<\eta <\eta _{c}^{\left( 2\right) }$ and $\eta >\eta
_{c}^{\left( 2\right) }$. All the phase transitions are of the first order,
due to the existence of the N-II-SR and SF-SR mixed phases.

\section{Parameter estimation and possible experimental observation}

\label{Parameter estimation}

We now take $^{40}$K atom as an example to estimate the related parameters.
For the fermionic $^{40}$K atoms with the Fermi energy $E_{F}\sim 0.73$ MHz,
the ground states with $^{2}$S$_{1/2}$ are given by $\left\vert \uparrow
\right\rangle =\left\vert F=9/2,m_{F}=9/2\right\rangle $ and $\left\vert
\downarrow \right\rangle =\left\vert F=9/2,m_{F}=7/2\right\rangle $, and the
excited states with $^{2}$P$_{1/2}$ are chosen as $\left\vert 1\right\rangle
=\left\vert F=9/2,m_{F}=7/2\right\rangle $ and $\left\vert 2\right\rangle
=\left\vert F=9/2,m_{F}=9/2\right\rangle $, where $F$ and $m_{F}$ denote the
total angular momentum and magnetic quantum numbers, respectively.

Due to the optical properties of the $^{40}$K D1-line, the cavity length and
the wavelengths of the transverse pumping lasers are chosen as $178$ $\mu $m
and $770$ nm, respectively. In this case, both the fermion-photon coupling
strengths $g_{1}$ and $g_{2}$ have the order of MHz, which is responsible
for the rotating-wave approximation for deriving the Hamiltonians (\ref{S3})
and (\ref{S4}). When the waist radius of the cavity mode is given by $27$ $%
\mu $m and the cavity has a finesse with the order of $10^{5}$, the cavity
decay rate $\kappa $ has the order of MHz. Since the effective
fermion-photon coupling strength $\eta =\sqrt{N}g_{1}\Omega _{1}^{\ast
}/\left( 2\Delta _{1}\right) $ $=\sqrt{N}g_{2}\Omega _{2}^{\ast }/\left(
2\Delta _{2}\right) $ is enhanced by a factor $\sqrt{N}$, it can reach the
order of MHz by varying the Rabi frequencies of the transverse pumping
lasers, even if the large detunings for ensuring the adiabatical
approximation in deriving the Hamiltonians (\ref{S9}) is taken into account.
The effective resonant frequency $\omega _{0}=$ $\left( \tilde{\omega}%
_{\downarrow }-\tilde{\omega}_{\uparrow }\right) /2$ and the atom-number
dependent cavity frequency $\omega =N\zeta +\tilde{\omega}$ are controlled
easily by varying the frequencies of the driving and transverse pumping
lasers. In experiments \cite{KB10}, $\omega _{0}$ and $\omega $ can be tuned
from $-$GHz to GHz, and even goes beyond this regime.

In addition, the 2D degenerate Fermi gas has been realized experimentally by
a 1D deep optical lattice along the third dimension, where the tunneling
between different layers is suppressed completely \cite%
{KM10,BF11,MF11,MK12,MGR15}. The 1D optical lattice potential $V_{0}\sin
^{2}(2\pi x/\lambda _{w})$\ can be generated using two counter-propagating
laser beams (parallel to the $x$\ axis with wavelength $\lambda _{w}$). In
such a case, $E_{b}\simeq 0.915\hbar \omega _{L}\exp (\sqrt{2\pi }%
l_{L}/a_{s})/\pi $, where $\omega _{L}=\sqrt{8\pi ^{2}V_{0}/(m\lambda
_{w}^{2})}$\ is the effective trapping frequency along the $x$\ axis, $l_{L}=%
\sqrt{\hbar /(m\omega _{L})}$, and $a_{s}$\ is the 3D s-wave scattering
length \cite{JL15}. Therefore, $E_{b}$\ can be tuned by varying the 3D
s-wave scattering length $a_{s}$\ via the Feshbash resonance and reach the
order of MHz \cite{CC10}. Based on the above estimation, all parameters used
to plot Figs.~\ref{figMP}-\ref{wphase08} could be realized in experiments.

Finally, we address briefly how to detect the predicted quantum phases and
phase diagrams, which are mainly governed by the mean-field gap $\Delta $,
the scaled mean-photon number $\left\vert \bar{\alpha}\right\vert ^{2}$, and
the scaled polarization $\bar{m}$. In experiments, the mean-field gap can be
measured by the radio-frequency excitation spectra, i.e., the fractional
loss of the fermions in one of the lowest substates through varying the
radio-frequency frequency \cite{CC04}, the polarization and the properties
of the mixed phase can be measured by observing the different density
distributions between the two-component Fermi gas \cite{GBP06,MWZ06}, and
the mean-photon number can be detected using calibrated single-photon
counting modules, which allow us to monitor the intracavity light intensity
\textit{in situ} \cite{KB10}. Based on these developed experimental
techniques, we believe that our predicted quantum phases and phase diagrams
could be detected in future experiments.\newline

\section{Discussion and Conclusion}

\label{Discussion}

Before ending up this paper, we make one remark. In real experiments, the
harmonic trap usually exists. For simplicity, here we have only considered a
weak harmonic trap that can be neglected. For more rigorous calculations, we
should apply the local density approximation \cite{WK65}, in which the
chemical potential becomes $\mu +V(r)$.

In summary, we have analytically investigated the ground-state properties of
a 2D polarized degenerate Fermi gas in a high-finesse optical cavity. By
solving the photon-number dependent BdG equation, we have found rich quantum
phases and phase diagrams, which depend crucially on the fermion-photon
coupling strength, the fermion-fermion interaction strength, and the atomic
resonant frequency (effective Zeeman field). In particular, without the
fermion-fermion interaction and with a weak atomic resonant frequency, we
have found a mixed phase that the N-II and SR phases coexist, and revealed a
first-order phase transition from the N-II phase to the SR phase. With the
intermediate fermion-fermion interaction and fermion-photon coupling
strengths, we have predicted another mixed phase that the SF and SR phases
coexist. Finally, we have presented a parameter estimation and have
addressed briefly how to detect these predicted quantum phases and phase
diagrams in experiments.\newline

\section{Acknowledgements}

This work is supported in part by the NSFC under Grants No.~11674200,
No.~11422433, No.~11604392, No.~11434007, and No.~61378049; the FANEDD under
Grant No.~201316; SFSSSP; OYTPSP; and SSCC. \newline

\appendix

\section{Derivation of Eqs.~(\protect\ref{NRG})-(\protect\ref{NRP})}

\label{AppAI}


When the scaled ground-state energies of the N-II and SR phases are equal,
these two phases coexist and the corresponding phase is called the N-II-SR
mixed phase. In order to fully describe the fundamental properties of this
mixed phase, we should introduce the fractions of the N-II and SR phases, $%
x_{0}$ and $1-x_{0}$, which are determined by \cite{LH08}
\begin{eqnarray}
n &=&x_{0}n_{N-II}\left( \mu _{N-II-SR}\text{, }\omega _{0}\right)  \notag \\
&&+\left( 1-x_{0}\right) n_{SR}\left( \mu _{N-II-SR}\text{, }\omega _{0}%
\text{, }\eta \right) ,  \label{FL}
\end{eqnarray}%
where $x_{0}$ $\in \left[ 0\text{, }1\right] $, $\mu _{N-II-SR}\in \left[
\mu _{N-II}\text{, }\mu _{SR}\right] $ is the chemical potential in the
N-II-SR mixed phase, and $n_{N-II}$ ($n_{SR}$) and $\mu _{N-II}$ ($\mu _{SR}$%
) are the atom density and the chemical potential in the N-II (SR) phase.
Notice that in contrast to the main text, in order to better analyze the
properties of the mixed phase, hereafter we make some marks of the different
quantum phases.\textbf{\ }When $x_{0}=1$, $\bar{E}_{\text{G}}^{\text{N-II}%
}\left( \mu _{N-II}\text{, }\omega _{0}\right) $, $\mu _{N-II}$, and $\bar{m}%
_{N-II}$ are governed by Eqs.~(\ref{GN2}) and (\ref{NTG}), and $%
n_{N-II}=n\mu _{N-II}/E_{F}$. When $x_{0}=0$, $\bar{E}_{\text{G}}^{\text{SR}%
}\left( \mu _{SR}\text{, }\omega _{0}\text{, }\eta \right) $, $\mu _{SR}$, $%
\left\vert \bar{\alpha}\right\vert _{SR}$, $n_{SR}$, and $\bar{m}_{SR}$ are
given by%
\begin{eqnarray}
\bar{E}_{\text{G}}^{\text{SR}} &=&-\frac{\mu _{SR}^{2}}{4E_{F}A}-\frac{%
\omega _{0}^{2}\left( \omega ^{2}+\kappa ^{2}\right) }{4\omega \eta ^{2}},
\label{GSR} \\
\mu _{SR} &=&2\left( E_{F}-\frac{\omega \eta ^{2}}{\omega ^{2}+\kappa ^{2}}%
\right) ,  \label{CPO} \\
\left\vert \bar{\alpha}\right\vert _{SR} &=&\frac{1}{2}\sqrt{\frac{\mu
_{SR}^{2}\eta ^{2}}{\left( \omega ^{2}+\kappa ^{2}\right) E_{F}^{2}A^{2}}-%
\frac{\omega _{0}^{2}\left( \omega ^{2}+\kappa ^{2}\right) }{\omega ^{2}\eta
^{2}}},  \label{SE} \\
n_{SR} &=&\frac{n\mu _{SR}}{2E_{F}A},  \label{SRM} \\
\bar{m}_{SR} &=&\frac{\omega _{0}\left( \omega ^{2}+\kappa ^{2}\right) }{%
4\omega \eta ^{2}}.  \label{SRP}
\end{eqnarray}

From the phase equilibrium condition $\bar{E}_{\text{G}}^{\text{N-II}}\left(
\mu _{N-II-SR}\text{, }\omega _{0}\right) =$ $\bar{E}_{\text{G}}^{\text{SR}%
}\left( \mu _{N-II-SR}\text{, }\omega _{0}\text{, }\eta \right) $ \cite%
{DES07,LH08} and Eqs.~(\ref{GN2}) and (\ref{GSR}) , when $\mu _{N-II-SR}=\mu
_{N-II}=E_{F}$, i.e., $x_{0}=1$, the phase boundary between the N-II phase
and the N-II-SR mixed phase is given by $\eta =\eta _{c}^{\left( 2\right) }$%
. When $\mu _{N-II-SR}=\mu _{SR}=2E_{F}-2\omega \eta ^{2}/\left( \omega
^{2}+\kappa ^{2}\right) $, i.e., $x_{0}=0$, the phase boundary between the
N-II-SR mixed phase and the SR phase is also given by $\eta =\eta
_{c}^{\left( 2\right) }$. In addition, in the N-II-SR mixed phase, $\mu
_{N-II-SR}$ is also derived from $\bar{E}_{\text{G}}^{\text{N-II}}\left( \mu
_{N-II-SR}\text{, }\omega _{0}\right) =$ $\bar{E}_{\text{G}}^{\text{SR}%
}\left( \mu _{N-II-SR}\text{, }\omega _{0}\text{, }\eta \right) $. The
result is given by
\begin{equation}
\mu _{N-II-SR}=E_{F},  \label{MM}
\end{equation}%
which is the same as Eq.~(\ref{NRM}). Substituting Eq.~(\ref{MM}), $n_{N-II}$%
, and $n_{SR}$ into Eq.~(\ref{FL}), we find $x_{0}$ and $1-x_{0}$ are
arbitrary values ranging from $0$ to $1$.

Finally, we prove that the ground state of the N-II-SR mixed phase is
stable. Due to existence of the fractions of the N-II and SR phases, the
scaled ground-state energy in this mixed phase is defined as \cite%
{DES07,LH08}
\begin{eqnarray}
&&\bar{E}_{\text{G}}^{\text{N-II-SR}}\left( \omega _{0}\text{, }\eta \right)
\notag \\
&=&\mu _{N-II-SR}+x_{0}\bar{E}_{\text{G}}^{\text{N-II}}\left( \mu _{N-II-SR}%
\text{, }\omega _{0}\right)  \notag \\
&&+\left( 1-x_{0}\right) \bar{E}_{\text{G}}^{\text{SR}}\left( \mu _{N-II-SR}%
\text{, }\omega _{0}\text{, }\eta \right) .  \label{ML}
\end{eqnarray}%
The differences between $\bar{E}_{\text{G}}^{\text{N-II-SR}}\left( \omega
_{0}\text{, }\eta \right) $ and $\bar{E}_{\text{G}}^{\text{SR}}\left( \omega
_{0}\right) $ or between $\bar{E}_{\text{G}}^{\text{N-II-SR}}\left( \omega
_{0}\text{, }\eta \right) $ and $\bar{E}_{\text{G}}^{\text{N-II}}\left(
\omega _{0}\right) $ are expressed as%
\begin{eqnarray}
\bar{E}_{\text{G}}^{\text{N-II-SR}}\left( \omega _{0}\text{, }\eta \right) -%
\bar{E}_{\text{G}}^{\text{N-II}}\left( \omega _{0}\right) &=&0,  \label{GFP}
\\
\bar{E}_{\text{G}}^{\text{N-II-SR}}\left( \omega _{0}\text{, }\eta \right) -%
\bar{E}_{\text{G}}^{\text{SR}}\left( \omega _{0}\right) &=&-\frac{%
E_{F}^{2}-\omega _{0}^{2}}{2E_{F}}.  \label{GSA}
\end{eqnarray}%
It can be seen clearly from Eqs.~(\ref{GFP}) and (\ref{GSA}) that these
energy differences are less than or equal to zero, i.e., the ground state of
the N-II-SR mixed phase is stable at $\eta =\eta _{c}^{\left( 2\right) }$.

Substituting Eqs.~(\ref{GN2}), (\ref{GSR}), and (\ref{MM}) into Eqs.~(\ref%
{SE}) and (\ref{ML}), we derive Eqs.~(\ref{NRG})-(\ref{NRA}). In addition,
according to Eq.~(\ref{FL}), we obtain Eq.~(\ref{NRP}).\newline

\section{Derivation of Eqs.~(\protect\ref{GCPP})-(\protect\ref{CPUU}) and (%
\protect\ref{GCP})-(\protect\ref{CPU})}

\label{AppAII}


\subsubsection{$\protect\omega _{0}\geqslant E_{F}$}

In the case of $\omega _{0}\geqslant E_{F}$, when the scaled ground-state
energies of the N-I and SF phases are equal, these two phases coexist and
the corresponding phase is called the N-I-SF mixed phase. In order to fully
describe the fundamental properties of this mixed phase, we introduce the
fractions of the N-I and SR phases, $x_{1}$ and $1-x_{1}$, which are
determined by \cite{LH08}
\begin{eqnarray}
n &=&x_{1}n_{SF}\left( \mu _{N-I-SF}\text{, }E_{b}\right)  \notag \\
&&+\left( 1-x_{1}\right) n_{N-I}\left( \mu _{N-I-SF}\text{, }\omega
_{0}\right) ,  \label{FST}
\end{eqnarray}%
where $x_{1}$ $\in \left[ 0\text{, }1\right] $, $\mu _{N-I-SF}\in \left[ \mu
_{SF}\text{, }\mu _{N-I}\right] $ is the chemical potential in the N-I-SF
mixed phase, and $n_{N-I}$ ($n_{SF}$) and $\mu _{N-I}$ ($\mu _{SF}$) are the
atom density and the chemical potential in the N-I (SF) phase. When $x_{1}=1$%
, we obtain
\begin{eqnarray}
\bar{E}_{\text{G}}^{\text{SF}} &=&-\frac{1}{2E_{F}}\left( \mu _{SF}+\frac{%
E_{b}}{2}\right) ^{2},  \label{GS} \\
\Delta _{SF} &=&\sqrt{E_{b}\left( E_{b}+2\mu _{SF}\right) },  \label{TD} \\
\mu _{SF} &=&E_{F}-\frac{E_{b}}{2},  \label{MP} \\
n_{SF} &=&\frac{n}{E_{F}}\left( \mu _{SF}+\frac{E_{b}}{2}\right) ,
\label{TN} \\
\text{ }m_{SF} &=&0.  \label{TP}
\end{eqnarray}%
When $x_{1}=0$, the corresponding $\bar{E}_{\text{G}}^{\text{N-I}}\left( \mu
_{N-I}\text{, }\omega _{0}\right) $, $\mu _{N-I}$, and $\bar{m}_{N-I}$ are
governed by Eq.~(\ref{GN1}) and (\ref{NIG}) and $n_{N-I}=\left( \omega
_{0}+\mu _{N-I}\right) /\left( 2E_{F}\right) $.

From the phase equilibrium condition $\bar{E}_{\text{G}}^{\text{N-I}}\left(
\mu _{N-I-SF}\text{, }\omega _{0}\right) =$ $\bar{E}_{\text{G}}^{\text{SF}%
}\left( \mu _{N-I-SF}\text{, }E_{b}\right) $ \cite{DES07,LH08} and Eqs.~(\ref%
{GN1}) and (\ref{GS}), when $\mu _{N-I-SF}=\mu _{SF}=E_{F}-E_{b}/2$, i.e., $%
x_{1}=1$, the phase boundary between the SF phase and the N-I-SF mixed phase
is given by $E_{b}=2\left[ \omega _{0}-\left( \sqrt{2}-1\right) E_{F}\right]
$. When $\mu _{N-I-SF}=\mu _{N-I}=2E_{F}-\omega _{0}$, i.e., $x_{1}=0$, the
phase boundary between the N-I-SF mixed phase and the N-I phase is given by $%
E_{b}=2\!\left[ \omega _{0}-\left( 2-\sqrt{2}\right) E_{F}\right] $. In
addition, in the N-I-SF mixed phase, $\mu _{N-I-SF}$ is also derived from $%
\bar{E}_{\text{G}}^{\text{N-I}}\left( \mu _{N-I-SF}\text{, }\omega
_{0}\right) =$ $\bar{E}_{\text{G}}^{\text{SF}}\left( \mu _{N-I-SF}\text{, }%
E_{b}\right) $. The result is given by%
\begin{equation}
\mu _{N-I-SF}=\frac{\sqrt{2}\omega _{0}-E_{b}}{2-\sqrt{2}},  \label{MI}
\end{equation}%
which is the same as Eq.~(\ref{CPMM}). Substituting Eqs.~(\ref{TN}), (\ref%
{MI}), and $n_{N-I}$ into Eq.~(\ref{FST}), we find
\begin{equation}
x_{1}=\frac{2\sqrt{2}E_{F}}{2\omega _{0}-E_{b}}-\sqrt{2}-1.  \label{XT}
\end{equation}

Finally, we prove that the N-I-SF mixed phase has a lowest ground-state
energy. Using Eqs.~(\ref{GS})-(\ref{TP}) and (\ref{MI})-(\ref{XT}), the
scaled ground-state energy in the N-I-SF mixed phase is defined as \cite%
{DES07,LH08}
\begin{eqnarray}
&&\bar{E}_{\text{G}}^{\text{N-I-SF}}\left( \omega _{0}\text{, }E_{b}\right)
\notag \\
&=&\mu _{N-I-SF}+x_{1}\bar{E}_{\text{G}}^{\text{SF}}\left( \mu _{N-I-SF}%
\text{, }E_{b}\right)  \notag \\
&&+\left( 1-x_{1}\right) \bar{E}_{\text{G}}^{\text{N-I}}\left( \mu _{N-I-SF}%
\text{, }\omega _{0}\right) .  \label{GMT}
\end{eqnarray}%
The differences between $\bar{E}_{\text{G}}^{\text{N-I-SF}}\left( \omega _{0}%
\text{, }E_{b}\right) $ and $\bar{E}_{\text{G}}^{\text{SF}}\left( \omega _{0}%
\text{, }E_{b}\right) $ or between $\bar{E}_{\text{G}}^{\text{N-I-SF}}\left(
\omega _{0}\text{, }E_{b}\right) $ and $\bar{E}_{\text{G}}^{\text{N-I}%
}\left( \omega _{0}\text{, }E_{b}\right) $ are expressed as%
\begin{eqnarray}
&&\bar{E}_{\text{G}}^{\text{N-I-SF}}\left( \omega _{0}\text{, }E_{b}\right) -%
\bar{E}_{\text{G}}^{\text{SF}}\left( \omega _{0}\text{, }E_{b}\right)  \notag
\\
&=&-\frac{\left( \sqrt{2}+1\right) ^{2}}{2E_{F}}\left( \omega
_{0}^{2}-E_{b}E_{F}^{2}+\frac{E_{b}^{2}}{4}\right) ^{2},  \label{GFS} \\
&&\bar{E}_{\text{G}}^{\text{N-I-SF}}\left( \omega _{0}\text{, }E_{b}\right) -%
\bar{E}_{\text{G}}^{\text{N-I}}\left( \omega _{0}\text{, }E_{b}\right)
\notag \\
&=&-\frac{\left( \sqrt{2}+1\right) ^{2}}{2E_{F}}\left( \omega
_{0}^{2}-E_{b}E_{F}^{2}-\frac{E_{b}^{2}}{4}\right) ^{2}.  \label{GFN}
\end{eqnarray}%
It can be seen clearly from Eqs.~(\ref{GFS}) and (\ref{GFN}) that these
energy differences are negative, i.e., the N-I-SF mixed phase has a lowest
scaled ground-state energy for $2\left[ \omega _{0}-\left( \sqrt{2}-1\right)
E_{F}\right] <E_{b}<2\!\left[ \omega _{0}-\left( 2-\sqrt{2}\right) E_{F}%
\right] $.

Substituting Eqs.~(\ref{GN1}), (\ref{GS}), (\ref{MI}), and (\ref{XT}) into
Eqs.~(\ref{TD}) and (\ref{GMT}), we derive Eqs.~(\ref{GCPP})-(\ref{CPMM}).
In addition, according to Eqs.~(\ref{FST}) and (\ref{XT}), we derive Eq.~(%
\ref{CPUU}).\newline

\subsubsection{$\protect\omega _{0}<E_{F}$}

In the case of $\omega _{0}<E_{F}$, when the scaled ground-state energies of
the N-II and SF phases are equal, these two phases coexist and the
corresponding phase is called the N-II-SF mixed phase. In order to fully
describe the fundamental properties of this mixed phase, we should introduce
the fractions of the N-II and SR phases, $x_{3}$ and $1-x_{3}$, which are
determined by \cite{LH08}
\begin{eqnarray}
n &=&x_{3}n_{SF}\left( \mu _{N-II-SF}\text{, }E_{b}\right)  \notag \\
&&+\left( 1-x_{3}\right) n_{N-II}\left( \mu _{N-II-SF}\text{, }\omega
_{0}\right) ,  \label{FS}
\end{eqnarray}%
where $x_{3}$ $\in \left[ 0\text{, }1\right] $ and $\mu _{N-II-SF}\in \left[
\mu _{SF}\text{, }\mu _{N-II}\right] $ is the chemical potential in the
N-II-SF mixed phase. When $x_{3}=1$, $E_{\text{G}}^{\text{SF}}\left( \mu
_{SF}\text{, }\omega _{0}\right) $, $\Delta _{SF}$, $\mu _{SF}$, $n_{SF}$,
and $m_{SF}$ are governed by Eqs.~(\ref{GS})-(\ref{TP}). When $x_{3}=0$, $%
\bar{E}_{\text{G}}^{\text{N-II}}\left( \mu _{N-II}\text{, }\omega
_{0}\right) $, $\mu _{N-II}$, and $\bar{m}_{N-II}$ are governed by Eqs.~(\ref%
{GN2}) and (\ref{NTG}), and $n_{N-II}=n\mu _{N-II}/E_{F}$.

From the phase equilibrium condition $\bar{E}_{\text{G}}^{\text{N-II}}\left(
\mu _{N-II-SF}\text{, }\omega _{0}\right) =$ $\bar{E}_{\text{G}}^{\text{SF}%
}\left( \mu _{N-II-SF}\text{, }E_{b}\right) $ \cite{DES07,LH08} and Eqs.~(%
\ref{NTG}) and (\ref{GS}), when $\mu _{N-II-SF}=\mu _{SF}=E_{F}-E_{b}/2$,
i.e., $x_{3}=1$, the phase boundary between the SF phase and the N-II-SF
mixed phase is given by $E_{b}=2\!\left( \!\sqrt{E_{F}^{2}+\omega _{0}^{2}}%
-E_{F}\right) $.\emph{\ }When $\mu _{N-II-SF}=\mu _{N-II}=E_{F}$, i.e., $%
x_{3}=0$, the phase boundary between the N-II-SF mixed phase and the N-II
phase is given by $E_{b}=2\left( \!E_{F}-\sqrt{E_{F}^{2}-\omega _{0}^{2}}%
\right) \Theta \left( \omega _{0}-\omega _{01}\right) $. In addition, in the
N-II-SF mixed phase, $\mu _{N-II-SF}$ is also derived from $\bar{E}_{\text{G}%
}^{\text{N-II}}\left( \mu _{N-II-SF}\text{, }\omega _{0}\right) =$ $\bar{E}_{%
\text{G}}^{\text{SF}}\left( \mu _{N-II-SF}\text{, }E_{b}\right) $. The
result is given by
\begin{equation}
\mu _{N-II-SF}=\frac{\omega _{0}^{2}}{E_{b}}-\frac{E_{b}}{4},  \label{UM}
\end{equation}%
which is the same as Eq.~(\ref{CPM}). Substituting Eqs.~(\ref{TN}), (\ref{UM}%
), and $n_{N-II}$ into Eq.~(\ref{FS}), we find
\begin{equation}
x_{3}=\frac{1}{2}+\frac{2E_{F}}{E_{b}}-\frac{2\omega _{0}^{2}}{E_{b}^{2}}.
\label{FI}
\end{equation}

Finally, we prove that the N-II-SF mixed phase has a lowest scaled
ground-state energy. Using Eqs.~(\ref{GS})-(\ref{FI}), the scaled
ground-state energy in this mixed phase is defined as \cite{DES07,LH08}
\begin{eqnarray}
&&\bar{E}_{\text{G}}^{\text{N-II-SF}}\left( \omega _{0}\text{, }E_{b}\right)
\notag \\
&=&\mu _{N-II-SF}+x_{3}\bar{E}_{\text{G}}^{\text{SF}}\left( \mu _{N-II-SF}%
\text{, }E_{b}\right)  \notag \\
&&+\left( 1-x_{3}\right) \bar{E}_{\text{G}}^{\text{N-II}}\left( \mu
_{N-II-SF}\text{, }\omega _{0}\right) .  \label{GM}
\end{eqnarray}%
The differences between $\bar{E}_{\text{G}}^{\text{N-II-SF}}\left( \omega
_{0}\text{, }E_{b}\right) $ and $\bar{E}_{\text{G}}^{\text{SF}}\left( \omega
_{0}\text{, }E_{b}\right) $ or between $\bar{E}_{\text{G}}^{\text{N-II-SF}%
}\left( \omega _{0}\text{, }E_{b}\right) $ and $\bar{E}_{\text{G}}^{\text{%
N-II}}\left( \omega _{0}\text{, }E_{b}\right) $ are expressed as%
\begin{eqnarray}
&&\bar{E}_{\text{G}}^{\text{N-II-SF}}\left( \omega _{0}\text{, }E_{b}\right)
-\bar{E}_{\text{G}}^{\text{SF}}\left( \omega _{0}\text{, }E_{b}\right)
\notag \\
&=&-\frac{1}{2E_{F}E_{b}^{2}}\left( \omega _{0}^{2}-E_{b}E_{F}^{2}+\frac{%
E_{b}^{2}}{4}\right) ^{2},  \label{DI} \\
&&\bar{E}_{\text{G}}^{\text{N-II-SF}}\left( \omega _{0}\text{, }E_{b}\right)
-\bar{E}_{\text{G}}^{\text{N-II}}\left( \omega _{0}\text{, }E_{b}\right)
\notag \\
&=&-\frac{1}{2E_{F}E_{b}^{2}}\left( \omega _{0}^{2}-E_{b}E_{F}^{2}-\frac{%
E_{b}^{2}}{4}\right) ^{2}.  \label{DII}
\end{eqnarray}%
It can be seen clearly from Eqs.~(\ref{DI}) and (\ref{DII}) that these
energy differences are negative, i.e., the N-II-SF mixed phase has a lowest
scaled ground-state energy for $2\!\left( \!\sqrt{E_{F}^{2}+\omega _{0}^{2}}%
-E_{F}\right) <E_{b}<2\left( \!E_{F}-\sqrt{E_{F}^{2}-\omega _{0}^{2}}\right)
\Theta \left( \omega _{0}-\omega _{01}\right) $.

Substituting Eqs.~(\ref{GN2}), (\ref{GS}), (\ref{UM}), and (\ref{FI}) into
Eqs.~(\ref{TD}) and (\ref{GM}), we derive Eqs.~(\ref{GCP})-(\ref{CPM}). In
addition, according to (\ref{FS}) and (\ref{FI}), we obtain Eq.~(\ref{CPU}).

\section{Derivation of Eqs.~(\protect\ref{GCP2})-(\protect\ref{CPV})}

\label{AppAIII}


When the scaled ground-state energies of the SF and SR phases are equal,
these two phases coexist and the corresponding phase is called the SF-SR
mixed phase. In order to fully describe the fundamental properties of this
mixed phase, we introduce the fractions of the SF and SR phases, $x_{2}$ and
$1-x_{2}$, which are determined by \cite{LH08}
\begin{eqnarray}
n &=&x_{2}n_{SF}\left( \mu _{SF-SR}\text{, }E_{b}\right)  \notag \\
&&+\left( 1-x_{2}\right) n_{SR}\left( \mu _{SF-SR}\text{, }\omega _{0}\text{%
, }\eta \right) ,  \label{FSI}
\end{eqnarray}%
where $x_{2}$ $\in \left[ 0\text{, }1\right] $ and $\mu _{SF-SR}\in \left[
\mu _{SF}\text{, }\mu _{SR}\right] $ is the chemical potential in the SF-SR
mixed phase. When $x_{2}=1$, $\bar{E}_{\text{G}}^{\text{SF}}\left( \mu _{SF}%
\text{, }E_{b}\right) $, $\Delta _{SF}$, $\mu _{SF}$, $n_{SF}$, and $m_{SF}$
are the same as the Eqs.~(\ref{GS})-(\ref{TN}). When $x_{2}=0$, $\bar{E}_{%
\text{G}}^{\text{SR}}\left( \mu _{SR}\text{, }\eta \right) $, $\left\vert
\alpha \right\vert _{SR}$, $\mu _{SR}$, $n_{SR}$, and $\bar{m}_{SR}$ are
governed by Eqs.~(\ref{GSR})-(\ref{SRP}).

From the phase equilibrium condition $\bar{E}_{\text{G}}^{\text{SF}}\left(
\mu _{SF-SR}\text{, }E_{b}\right) =$ $\bar{E}_{\text{G}}^{\text{SR}}\left(
\mu _{SF-SR}\text{, }\eta \right) $ \cite{DES07,LH08} and Eqs.~(\ref{GSR})
and (\ref{GS}), when $\mu _{SF-SR}=\mu _{SF}=E_{F}-E_{b}/2$, i.e., $x_{2}=1$%
, the phase boundary between the SF phase and the SF-SR mixed phase is given
by $E_{b}=E_{b}^{\left( 2\right) }\left( \eta -\eta _{c}^{\left( 3\right)
}\right) $.\emph{\ }When $\mu _{SF-SR}=\mu _{SR}=2E_{F}-2\omega \eta
^{2}/\left( \omega ^{2}+\kappa ^{2}\right) $, i.e., $x_{2}=0$, the phase
boundary between the SF-SR mixed phase and the SR phase is given by $%
E_{b}=E_{b}^{\left( 1\right) }$. In addition, in the SF-SR mixed phase, $\mu
_{SF-SR}$ is also derived from $\bar{E}_{\text{G}}^{\text{SF}}\left( \mu
_{SF-SR}\text{, }E_{b}\right) =$ $\bar{E}_{\text{G}}^{\text{SR}}\left( \mu
_{SF-SR}\text{, }\eta \right) $. The result is given by
\begin{equation}
\mu _{SF-SR}=\frac{-E_{b}A\pm \sqrt{A^{2}E_{b}^{2}-2A\left( 2A-1\right) B}}{%
2A-1},  \label{SRC}
\end{equation}%
which is the same as Eq.~(\ref{VPC}). Using Eqs.~(\ref{FSI}), (\ref{TN}), (%
\ref{SRM}), and (\ref{SRC}), $x_{2}$ is determined by
\begin{equation}
x_{2}=\frac{2E_{F}A-\mu _{SF-SR}}{2\mu _{SF-SR}A-\mu _{SF-SR}+E_{b}A}.
\label{XII}
\end{equation}

Finally, we prove that the SF-SR mixed phase has a lowest scaled
ground-state energy. Using Eqs.~(\ref{GSR})-(\ref{SRP}), (\ref{GS})-(\ref{TP}%
), and (\ref{SRC})-(\ref{XII}), The scaled ground-state energy in the SF-SR
mixed phase is defined as \cite{DES07,LH08}
\begin{eqnarray}
&&\bar{E}_{\text{G}}^{\text{SF-SR}}\left( \omega _{0}\text{, }E_{b}\text{, }%
\eta \right)  \notag \\
&=&\mu _{SF-SR}+x_{2}\bar{E}_{\text{G}}^{\text{SF}}\left( \mu _{SF-SR}\text{%
, }E_{b}\right)  \notag \\
&&+\left( 1-x_{2}\right) \bar{E}_{\text{G}}^{\text{SR}}\left( \mu _{SF-SR}%
\text{, }\eta \right) .  \label{GFI}
\end{eqnarray}%
The differences between $\bar{E}_{\text{G}}^{\text{SF-SR}}\left( \omega _{0}%
\text{, }E_{b}\text{, }\eta \right) $\ and $\bar{E}_{\text{G}}^{\text{SF}%
}\left( \omega _{0}\text{, }E_{b}\text{, }\eta \right) $\ or between $\bar{E}%
_{\text{G}}^{\text{SF-SR}}\left( \omega _{0}\text{, }E_{b}\text{, }\eta
\right) $\ and $\bar{E}_{\text{G}}^{\text{SR}}\left( \omega _{0}\text{, }%
E_{b}\text{, }\eta \right) $\ are given by%
\begin{eqnarray}
&&\bar{E}_{\text{G}}^{\text{SF-SR}}\left( \omega _{0}\text{, }E_{b}\text{, }%
\eta \right) -\bar{E}_{\text{G}}^{\text{SF}}\left( \omega _{0}\text{, }E_{b}%
\text{, }\eta \right)  \notag \\
&=&-\frac{1}{2E_{F}}\left( \frac{\mu _{M}^{2}}{2A}+E_{F}^{2}-E_{F}E_{b}%
\right)  \notag \\
&&+\mu _{M}-\frac{\omega _{0}^{2}\left( \omega ^{2}+\kappa ^{2}\right) }{%
4\omega \eta ^{2}},  \label{GFF} \\
&&\bar{E}_{\text{G}}^{\text{SF-SR}}\left( \omega _{0}\text{, }E_{b}\text{, }%
\eta \right) -\bar{E}_{\text{G}}^{\text{SR}}\left( \omega _{0}\text{, }E_{b}%
\text{, }\eta \right)  \notag \\
&=&\mu _{M}-E_{F}-\frac{\mu _{M}^{2}}{4E_{F}A}+\frac{\omega \eta ^{2}}{%
\omega ^{2}+\kappa ^{2}}.  \label{GFR}
\end{eqnarray}%
It can be seen clearly from Eqs.~(\ref{GFF}) and (\ref{GFR}) that these
energy differences are negative, i.e., the N-II-SF mixed phase has a lowest
scaled ground-state energy for $\!E_{b}^{\left( 1\right)
}<E_{b}<E_{b}^{\left( 2\right) }\left( \eta -\eta _{c}^{\left( 3\right)
}\right) $.

Substituting Eqs.~(\ref{GSR}), (\ref{GS}), (\ref{SRC}), and (\ref{XII}) into
Eqs.~(\ref{SE}) and (\ref{GFI}), we derive Eqs.~(\ref{GCP2})-(\ref{CPA}). In
addition, according to Eqs.~(\ref{FSI}) and (\ref{XII}), we obtain Eq.~(\ref%
{CPV}).\newline

\end{document}